\documentclass[pre,11pt,onecolumn,notitlepage,nofootinbib]{revtex4-1}
\pdfoutput = 1

\usepackage{amsmath,amssymb,graphicx,tabularx,diagbox,bm,chemarr}

\usepackage{color}
\definecolor{darkblue}{rgb}{0,0,0.6}
\definecolor{darkred}{rgb}{0.6,0,0}
\definecolor{darkgreen}{rgb}{0,0.6,0}

\usepackage[colorlinks=true,urlcolor=darkblue,citecolor=darkblue,linkcolor=darkblue,hyperfootnotes=false]{hyperref}

\newcommand{\rL}{\mathrm{L}}
\newcommand{\rR}{\mathrm{R}}
\newcommand{\rc}{\mathrm{c}}
\newcommand{\rs}{\mathrm{s}}
\newcommand{\bn}{\bar{n}}
\newcommand{\hrho}{\hat{\rho}}

\newcommand{\rexp}{\mathrm{e}}
\newcommand{\bU}{\bar{U}}

\newcommand{\hn}{\hat{n}}
\newcommand{\bv}{\bar{v}}
\newcommand{\hphi}{\hat\varphi}

\begin{document}

\title{Finite-size and finite-time effects in large deviation functions\\near dynamical symmetry breaking transitions}
\author{Yongjoo Baek}
\affiliation{DAMTP, Centre for Mathematical Sciences, University of Cambridge, Cambridge CB3 0WA, United Kingdom}
\altaffiliation[Current address: ]{Department of Physics and Astronomy, Seoul National University, Seoul 06977, Korea}
\email{y.baek@snu.ac.kr}

\author{Yariv Kafri}
\affiliation{Department of Physics, Technion, Haifa 32000, Israel}

\author{Vivien Lecomte}
\affiliation{Université Grenoble Alpes, CNRS, LIPhy, 38000 Grenoble, France}

\begin{abstract}
We introduce and study a class of particle hopping models consisting of a single box coupled to a pair of reservoirs. Despite being zero-dimensional, in the limit of large particle number and long observation time, the current and activity large deviation functions of the models can exhibit symmetry-breaking dynamical phase transitions. We characterize exactly the critical properties of these transitions, showing them to be direct analogues of previously studied phase transitions in extended systems. The simplicity of the model allows us to study features of dynamical phase transitions which are not readily accessible for extended systems. In particular, we quantify finite-size and finite-time scaling exponents using both numerical and theoretical arguments. Importantly, we identify an analogue of critical slowing near symmetry breaking transitions and suggest how this can be used in the numerical studies of large deviations. All of our results are also expected to hold for extended systems.
%
%We propose a class of particle hopping models consisting of a single-box system coupled to a pair of reservoirs. Despite being zero-dimensional, current or activity large deviations of the models exhibit symmetry-breaking dynamical phase transitions analogous to those of extended diffusive systems driven by boundary reservoirs. In the joint limit of infinitely many particles and a long observation period, the critical properties near the transitions are  described by theories derived from first principles. The simplicity of the model allows us to identify finite-size and finite-time scaling exponents by both numerical diagonalization and theoretical arguments. Our results clarify the essential elements of dynamical symmetry breaking and provide a guideline for numerical analysis of the associated critical phenomena.
\end{abstract}

\maketitle

\tableofcontents

\section{Introduction}

%low-noise limit

In recent years, there has been much interest in large deviation functions (LDFs, see~\cite{TouchettePR2009} for a review) encoding the probability of atypical fluctuations in time-averaged observables of many-body quantum~\cite{LevitovJETP1993,LevitovJMP1996,PilgramPRL2003,JordanJMP2004,Derezinski2008,EspositoRMP2009,ZnidaricPRL2014,GenwayJPA2014,CarolloPRE2017} and classical stochastic systems~\cite{DerridaJSM2007,DerridaJSP1999,DerridaJSP2004,BodineauPRL2004,Maes_2008,ProlhacJPA2008,BodineauJSP2008,ImparatoPRE2009,LecomtePTPS2010,PradosPRL2011,DeGierPRL2011,LazarescuJPA2011,DerridaJSM2011,GorissenPRL2012,GorissenPRE2012,KrapivskyPRE2012,FlindtPRL2013,AkkermansEPL2013,MeersonJSM2013,*MeersonPRE2014,HurtadoJSP2014,LazarescuJPA2015,ZarfatyJSM2016,KlymkoPRE2017,KlymkoPRE2018,WhitelamPRE2018}. Of special interest have been LDFs of the time-averaged current and activity, the latter quantifying the mean frequency of dynamical events during a given observation period. Since both quantities are determined by the full history rather than the instantaneous state, even in thermal equilibrium, their LDFs can exhibit unexpected behaviors. In particular, even if the steady-state probability distribution of instantaneous quantities, such as the density profile of particles in the system, contains no singularities, the LDF of {\it time-averaged quantities} can be singular, giving rise to a dynamical phase transition (DPT). This happens since the dominant history leading to a given atypical time-averaged quantity can change in an abrupt way as the value of the time-averaged quantity is varied. Like equilibrium phase transitions, DPTs can occur as first, second, or even higher-order singularities of LDFs. To date, DPTs have been found in a host of systems encompassing driven diffusive systems~\cite{HarrisJSM2005,BertiniPRL2005,BertiniJSP2006,BodineauPRE2005,BodineauCRP2007,AppertRollandPRE2008,ProlhacJPA2009,HurtadoPRL2009,*HurtadoPRL2011,LecomteJPA2012,HirschbergJSM2015,JackPRL2015,BaekPRL2017,BaekJPA2018,ShpielbergPRE2017a,ShpielbergPRE2017b,ShpielbergPRE2018}, kinetically constrained models~\cite{GarrahanPRL2007,GarrahanJPA2009,BodineauJSP2012,NemotoJSM2014,NemotoPRL2017}, interface growth~\cite{MajumdarJSM2014,LeDoussalEPL2016,JanasPRE2016,SmithPRE2018}, and active particles~\cite{CagnettaPRL2017,NemotoPRE2019}.

Most of the DPTs have been obtained in many-body extended systems\footnote{See~\cite{SpeckEPL2007,TsobgniNyawoEPL2016,*TsobgniNyawoPRE2018,GarrahanJPA2009,VaikuntanathanPRE2014,GingrichPRE2014} for exceptions.} whose sizes are taken to be infinite. It is natural to ask how much of the observed phenomenology is related to the fact that these systems are extended. In this paper, we address this question by introducing a class of models consisting of a one-site (or {\em single-box}) system connected to a pair of reservoirs and studying their current and activity large deviations. Instead of taking a limit where the system size goes to infinity, we utilize a recently introduced formalism~\cite{BaekJSM2016} where $N$, the maximum number of particles in the box, is arbitrarily large. Applying the saddle-point method, it is shown that even such models can exhibit DPTs induced by the breaking of the particle-hole symmetry, which was theoretically predicted~\cite{BaekPRL2017,BaekJPA2018} and numerically observed~\cite{PerezEspigaresPRE2018} in extended systems, with exactly the same critical exponents.

Importantly, the reduced dimensionality of a single-box model allows us to easily predict and confirm the effects of finite time, $T$, and finite size, $N$, on the critical phenomena near a symmetry-breaking DPT for arbitrary hopping rates. In previous studies of extended systems, finite-size scaling theories have been proposed for first and second-order DPTs of an exclusion process ~\cite{AppertRollandPRE2008,ShpielbergPRE2018,GorissenJPA2011} as well as for kinetically constrained models~\cite{Bodineau2012,BodineauJSP2012,NemotoJSM2014,NemotoPRL2017}. Much less is known about finite-time effects\footnote{%
  As we will see, the LDF in the infinite-time limit is given by the maximum eigenvalue of a well-defined operator, while the finite-time behavior of the LDF involves more eigenvalues.%
},
with only a few results concerning diffusive~\cite{KrapivskyPRL2014} and super-diffusive~\cite{ProlhacPRL2016} relaxations of density fluctuations far away from any DPTs. For symmetry-breaking DPTs in extended systems with open boundaries, Ref.~\cite{BaekJPA2018} used heuristic arguments to predict finite-time and finite-size scaling exponents. These, however, have not been verified. In this paper, based on studies of finite-$T$ saddle-point trajectories and an exact diagonalization of the transition matrix at finite $N$, we identify both the finite-$T$ and finite-$N$ scaling exponents and propose a scaling form encompassing both. In particular, we are able to characterize in detail the different finite-$T$ scaling regimes. We find a regime where the initial condition strongly influence the LDF and, as one might expect, a late regime where the initial conditions do not play any role.
The results show that, near a symmetry-breaking DPT, a phenomenon analogous to critical slowing appears. Namely, the relaxation of the system from a given initial condition becomes anomalously slow as the DPT is approached. This might be used to locate such DPTs in numerics~\cite{GiardinaPRL2006,LecomteJSM2007,TailleurModeling2009,GiardinaJSP2011,NemotoPRL2014,NemotoPRE2016,RayPRL2018,BrewerJSM2018,PerezEspigaresArXiv2019} and possibly experiments by data collapse.

The paper is organized as follows. In Sec.~\ref{sec:model}, we introduce the single-box models and present a path-integral representation of their statistics. In Sec.~\ref{sec:symbreak}, we discuss how the theory of symmetry-breaking DPTs and the associated critical behaviors can be derived using a saddle-point method in the joint limit $T\to\infty$ and $N\to\infty$. In Sec.~\ref{sec:scaling_theory}, based on both numerical diagonalization and theoretical arguments, we study finite-size and finite-time effects, allowing us to characterize the critical features of the DPT. Finally, we conclude in Sec.~\ref{sec:conclusions}.

%Sec.~\ref{sec:model}, \ref{sec:symbreak}, \ref{sec:scaling_theory} \ref{sec:conclusions}

\section{Single-box models with particle-hole symmetry}
\label{sec:model}

In this section, we describe the general setup considered in our study. First, we introduce a general class of single-box models. Focusing on a subclass of such systems which obey a particle-hole symmetry, we formulate their coarse-grained descriptions for large $N$. This allows us to study their DPTs using saddle-point asymptotics.

\subsection{General single-box models}

\begin{figure}
\includegraphics[width=0.7\columnwidth]{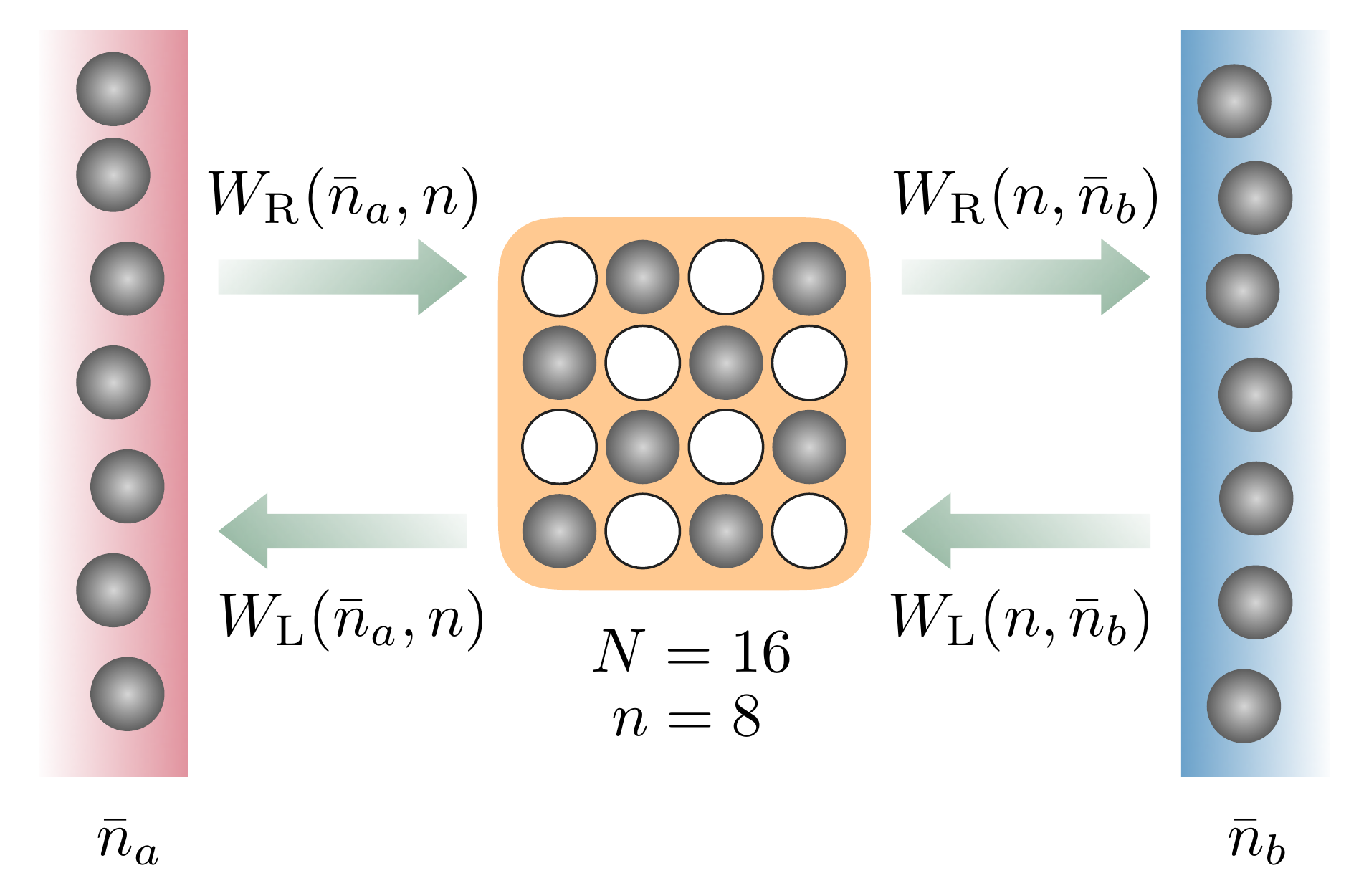}
\caption{\label{fig:fig1} Illustration of a generic single-box model. The hopping rates are determined by $n$, the number of particles in the box, and $\bar n_a$ ($\bar n_b$), the number of particles imposed by the left (right) reservoir. In the example shown here, the box holds $n = 8$ particles, while it can store at most $N = 16$ particles.}	
\end{figure}

We consider a single box, whose state is characterized by the number of particles $n$ inside. The box can hold at most $N$ particles ($0 \le n \le N$) and is coupled to a pair of particle reservoirs. The left (right) reservoir is described as a box with a fixed number of particles $\bn_a$ ($\bn_b$). The particles are exchanged with the left reservoir according to
\begin{align} \label{eq:left_res}
n \xrightleftharpoons[W_\rL(\bn_a,n+1)]{W_\rR(\bn_a,n)} n+1,
\end{align}
where $W_\rR(n_1,n_2)$ ($W_\rL(n_1,n_2)$) denotes the rate of hopping from the left (right) box to the right (left), see Fig.~\ref{fig:fig1}. Similarly, the exchange with the right reservoir is described by
\begin{align} \label{eq:right_res}
n \xrightleftharpoons[W_\rR(n+1,\bn_b)]{W_\rL(n,\bn_b)} n+1.
\end{align}

We are interested in the statistics of current and activity during a time interval $t \in [0,T]$. Defining the number $M_\rR(T)$ ($M_\rL(T)$) of rightward (leftward) hops across any of the two bonds connecting the reservoirs to the system, we have the time-averaged current per bond
\begin{align} \label{eq:j_def}
J_T \equiv \frac{1}{2T}\left[M_\rR(T) - M_\rL(T)\right]
\end{align}
and the time-averaged activity per bond
\begin{align} \label{eq:k_def}
K_T \equiv \frac{1}{2T}\left[M_\rR(T) + M_\rL(T)\right].
\end{align}
The joint scaled cumulant generating function (CGF) $\Psi(\lambda,\mu)$ for $J_T$ and $K_T$ is defined as
\begin{align} \label{eq:Psi_def}
\Psi(\lambda,\mu) \equiv \lim_{T\to\infty} \frac{1}{T} \ln \left\langle \rexp^{T(\lambda J_T + \mu K_T)}\right\rangle,
\end{align}
where $\langle\cdot\rangle$ denotes the average over histories. Using standard methods, described in Appendix~\ref{app:path}, one can show that
\begin{align} \label{eq:cgf_path_integ}
\rexp^{T \Psi(\lambda,\mu)} = \left\langle \rexp^{T(\lambda J_T + \mu K_T)}\right\rangle = \int \mathcal{D}[n,\hn]\,\rexp^{-\int_0^T dt\, \left[\hn\dot{n} - \mathcal{H}_{\lambda,\mu}(n,\hn)\right]},
\end{align}
with an effective Hamiltonian
\begin{align} \label{eq:h_before_res_0}
\mathcal{H}_{\lambda,\mu}(n,\hn) &\equiv W_\rR(\bn_a,n)\left[\rexp^{\hn+(\mu+\lambda)/2}-1\right] + W_\rL(\bn_a,n)\left[\rexp^{-\hn+(\mu-\lambda)/2}-1\right] \nonumber\\
&\quad + W_\rL(n,\bn_b)\left[\rexp^{\hn+(\mu-\lambda)/2}-1\right] + W_\rR(n,\bn_b)\left[\rexp^{-\hn+(\mu+\lambda)/2}-1\right].
\end{align}
Here $\hn$ is a momentum (integrated along the imaginary axis) conjugate to $n$, and the Lagrange multiplier $\lambda$ ($\mu$) is a counting variable conjugate to $J_T$ ($K_T$).

We are mainly interested in models presenting second-order singularities in the scaled CGF. As we show below, these naturally occur for a class of models whose dynamics obey a particle-hole symmetry. For simplicity, we first consider the case where the two reservoirs have equal densities $\bn_a = \bn_b = \frac{N}{2}$, which captures all the essential physics of the DPT. The generalization to the boundary-driven case $\bn_a \neq \bn_b$ is discussed in Appendix~\ref{app:boundary_driving}.

\subsection{Particle-hole symmetric models}

The particle-hole symmetry is implemented by choosing a dynamics which is invariant under the combined operation of the particle-hole exchange and the exchange of the reservoir locations. This is achieved by imposing
\begin{align} \label{eq:p-h_sym_rates}
W_\rR(n_1,n_2) = W_\rR(N-n_2,N-n_1),\quad
W_\rL(n_1,n_2) = W_\rL(N-n_2,N-n_1).
\end{align}
As stated above, we focus on the case where the reservoir densities are $N/2$.
We also assume that each hopping across a bond obeys local detailed balance, so that the rate of a rightward hop and that of a leftward one differ only due to a global field (bulk drive):
\begin{align} \label{eq:nu}
\frac{W_\rR(n_1,n_2)}{W_\rL(n_2,n_1)} = \nu.
\end{align}
Here $\nu > 0$ controls the strength of the field. To simplify the notation, we write the rate of a rightward hop from the left reservoir into the box as 
\begin{align} \label{eq:factorize}
W_\rR\!\left(\frac{N}{2},n\right) = \alpha\,V(n).
\end{align}
Then, using Eqs.~\eqref{eq:p-h_sym_rates}, \eqref{eq:nu}, and \eqref{eq:factorize}, the four hopping rates in Eqs.~\eqref{eq:left_res} and \eqref{eq:right_res} can be written as
\begin{align} \label{eq:rates_simplify}
W_\rR\!\left(\frac{N}{2},n\right) &= \alpha\,V(n),&
W_\rL\!\left(\frac{N}{2},n\right) &= \frac{\alpha}{\nu}\,V(N-n), \nonumber\\
W_\rL\!\left(n,\frac{N}{2}\right) &= \frac{\alpha}{\nu}\,V(n),&
W_\rR\!\left(n,\frac{N}{2}\right) &= \alpha\,V(N-n).
\end{align}
We note that, to impose the bound $0 \le n \le N$, the hopping rates are further constrained by
\begin{align} \label{eq:rate_n_constraint}
V(N) = 0.
\end{align}
With these choices, the Hamiltonian in Eq.~\eqref{eq:h_before_res_0} takes the form
\begin{align} 
\mathcal{H}_{\lambda,\mu}(n,\hn)
&= \alpha\left[\frac{\nu+1}{\nu}\left(\rexp^{\hn+\mu/2}\cosh\frac{\lambda}{2}-1\right)+\frac{\nu-1}{\nu}\rexp^{\hn+\mu/2}\sinh\frac{\lambda}{2}\right]V(n)\nonumber\\
&\quad + \alpha\left[\frac{\nu+1}{\nu}\left(\rexp^{-\hn+\mu/2}\cosh\frac{\lambda}{2}-1\right)+\frac{\nu-1}{\nu}\rexp^{-\hn+\mu/2}\sinh\frac{\lambda}{2}\right]V(N-n),
\end{align}
which can be rewritten as
\begin{align} \label{eq:h_before_res}
\mathcal{H}_{\lambda,\mu}(n,\hn) &= \gamma\left[\left(z\rexp^{\hn}-1\right)V(n)+\left(z\rexp^{-\hn}-1\right)V(N-n)\right].
\end{align}
Here we used definitions $\gamma \equiv (\nu+1)/\nu$ and
\begin{align} \label{eq:z_def}
z(\lambda,\mu) \equiv \frac{\rexp^{\mu/2}\cosh\!\left(\frac{\lambda}{2}+\tanh^{-1}\frac{\nu-1}{\nu+1}\right)}{\cosh\!\left(\tanh^{-1}\frac{\nu-1}{\nu+1}\right)}.
\end{align}
We note that the unbiased state $\lambda = \mu = 0$ corresponds to $z = 1$. From Eqs.~\eqref{eq:cgf_path_integ}, \eqref{eq:h_before_res}, and \eqref{eq:z_def}, one observes that the scaled CGF $\Psi$ depends on $\lambda$ and $\mu$ only through $z$. We also note that $z$ satisfies
\begin{align} \label{eq:gc_sym}
z(\lambda,\mu) = z\!\left(-\lambda - 4\tanh^{-1}\frac{\nu-1}{\nu+1},\mu\right),
\end{align}
which reflects the Gallavotti--Cohen symmetry~\cite{GallavottiPRL1995,*GallavottiJSP1995}.

So far we have described the {\em microscopic} dynamics in the sense that the discrete nature of the particles is maintained. We next formulate a coarse-grained description of the dynamics for large $N$, which makes the models easier to study by changing to continuous state variables and facilitating saddle-point techniques.

\subsection{Coarse-grained description for large $N$}

To take the large-$N$ limit, it is useful to define the rescaled fields $(\rho,\hrho)$ and introduce the rescaled time $t$ and observables 
\begin{align} \label{eq:rescaling}
n &\to N\rho, & \hn &\to \hrho, & V(n) &\to N^{k} v(\rho), \nonumber\\
t &\to N^{1-k}t,
& J_T &\to N^k J_T, & K_T &\to N^k K_T,
\end{align}
where $k$ is a positive number determined by the structure of the hopping rates (see below for examples). We note that the constraint~\eqref{eq:rate_n_constraint} can now be written as
\begin{align} \label{eq:rate_r_constraint}
v(1) = 0,
\end{align}
which ensures $0 \le \rho \le 1$. Using these in Eqs.~\eqref{eq:cgf_path_integ} and \eqref{eq:h_before_res}, we obtain a rescaled path-integral representation for the scaled CGF $\psi(z(\lambda,\mu)) = N^{-k}\Psi(\lambda,\mu)$, namely
\begin{align} \label{eq:path_integ_r}
\rexp^{N T \psi(z)} = \left\langle \rexp^{NT(\lambda J_T + \mu K_T)}\right\rangle = \int \mathcal{D}[\rho,\hrho]\,\rexp^{-NS_z[\rho,\hat{\rho}]}
\end{align}
with the action
\begin{align} \label{eq:action_r}
S_z[\rho,\hrho] \equiv \int_0^T dt\left[\hat{\rho}\dot{\rho} - H_z(\rho,\hat{\rho})\right],
\end{align}
where the Hamiltonian is given by
\begin{align} \label{eq:hamil_r}
H_z(\rho,\hrho) &\equiv \gamma\left[\left(z\rexp^{\hrho}-1\right)v(\rho)+\left(z\rexp^{-\hrho}-1\right)v(1-\rho)\right].
\end{align}
The particle-hole symmetry of the system is reflected in the symmetry of the action
\begin{align} \label{eq:p-h_sym_action}
S_z[\rho,\hrho] = S_z[1-\rho,-\hrho] \;.
\end{align}
For $N \gg 1$, from Eqs.~\eqref{eq:path_integ_r}, \eqref{eq:action_r}, and \eqref{eq:hamil_r}, we find that $\psi$ can be obtained by a saddle-point asymptotics
\begin{align} \label{eq:psi_saddle_approx}
\psi(z) = -\inf_{\rho,\hrho}\lim_{T\to\infty}\frac{1}{T} S_z[\rho,\hrho],
\end{align}
where the minimum action is achieved by real-valued $\rho$ and $\hrho$ obeying the Hamiltonian dynamics
\begin{align}
\dot{\rho} = \frac{\partial H_z}{\partial \hrho}
&= \gamma z\left[v(\rho)\,\rexp^{\hrho} - v(1-\rho)\,\rexp^{-\hrho}\right], \label{eq:drdt}\\
\dot{\hrho} = -\frac{\partial H_z}{\partial \rho}
&= -\gamma \left[v'(\rho)\,(z\,\rexp^{\hrho}-1)-v'(1-\rho)\,(z\,\rexp^{-\hrho}-1)\right]. \label{eq:drhdt}
\end{align}
Although $\psi(z)$ is defined only in the $T\to\infty$ limit, the above saddle-point trajectories still describe the histories dominantly contributing to the finite-time scaled CGF
\begin{align}
\psi_T(z) = -\inf_{\rho,\hrho}\frac{1}{T} S_z[\rho,\hrho]
\end{align}
whenever $N$ is large.

\section{Symmetry-breaking dynamical phase transitions}
\label{sec:symbreak}

We now calculate the scaled CGF $\psi$ of the single-box model and show that, with a proper choice of rates, the model displays the same DPTs exhibited by extended systems. In particular, we are interested in the DPTs between a particle-hole symmetric phase and one where the symmetry is broken.

\subsection{Particle-hole symmetric phase}

It is easy to see that, for any $\lambda$ and $\mu$,
\begin{align} \label{eq:sym_saddle}
\rho = 1/2, \quad \hrho = 0
\end{align}
yields a time-independent, particle-hole symmetric solution for Eqs.~\eqref{eq:drdt} and \eqref{eq:drhdt}. If this symmetric saddle-point profile truly minimizes the action, Eq.~\eqref{eq:psi_saddle_approx} implies
\begin{align} \label{eq:phi_sym_def}
\psi(z) = \psi^\text{sym}(z) &\equiv -\lim_{T\to\infty}\frac{1}{T}\,S_z[\rho(t) = 1/2,\hrho(t) = 0] = 2\gamma\bv (z-1).
\end{align}
Note that from here on we use the shorthand notations
\begin{align}
\bv = v(1/2),\,\bv' = v'(1/2),\, \bv'' = v''(1/2),\, \bv^{(n)} = v^{(n)}(1/2) \text{ for $n \ge 3$}.
\end{align}
In Appendix~\ref{sec:ssd}, we discuss the condition for the symmetric solution in Eq.~\eqref{eq:sym_saddle} to be the dominant profile in the unbiased state $z = 1$. We find that $v(\rho)$ being a monotonically decreasing function of $\rho$ is a sufficient condition. We also note that the mean current and activity are obtained from the above relations as
\begin{align}
\langle J \rangle = \left.\partial_\lambda \psi(\lambda,\mu)\right|_{\lambda=\mu=0} = \gamma\bv\,\frac{\nu-1}{\nu+1}, \quad
\langle K \rangle = \left.\partial_\mu \psi(\lambda,\mu)\right|_{\lambda=\mu=0} = \gamma \bv.
\end{align}
A second-order DPT occurs when this symmetric solution becomes unstable with respect to small fluctuations as the value of $z$ is changed. To this end, in the next section we study the Gaussian fluctuations of the action.

\subsection{Stability analysis}
\label{ssec:stability_analysis}

The fluctuations of the action around the symmetric saddle-point solution~\eqref{eq:sym_saddle}, 
\begin{align} \label{eq:S_exp}
S_z[1/2+\varphi(t),i\hphi(t)] = S_z[1/2,0] + \delta^2 S_z[\varphi,\hphi] + O\!\left(\varphi^3,\hphi\varphi^2,\hphi^2\varphi,\hphi^3\right),
\end{align}
are described by the Gaussian action
\begin{align} \label{eq:d2S}
\delta^2 S_z[\varphi,\hphi] &= \int dt\, \left[i\hphi\,\partial_t\varphi + \gamma z\bv\hphi^2 - 2i \gamma z\bv'\hphi\varphi +\gamma(1-z)\bv''\varphi^2\right] \nonumber\\
&= \gamma \int \frac{d\omega}{2\pi} \begin{bmatrix}
 \varphi_\omega & \hphi_\omega	
 \end{bmatrix}
\underbrace{
\begin{bmatrix}
2(1-z)\bv'' & -\frac{\omega+4iz\bv'}{2} \\
\frac{\omega-4iz\bv'}{2} & 2z\bv
\end{bmatrix}
}_{\mathbb{M}}
\begin{bmatrix}
\varphi_{-\omega} \\
\hphi_{-\omega}
\end{bmatrix},
\end{align}
where $\varphi_\omega$ and $\hphi_\omega$ are Fourier transforms of $\varphi$ and $\hphi$ defined as
\begin{align}
\varphi_\omega \equiv \int dt \,\varphi(t)\,\mathrm{e}^{-i\omega t},
\quad \hphi_\omega \equiv \int dt \,\hphi(t)\,\mathrm{e}^{-i\omega t} \;.
\end{align}
The eigenvalues of $\mathbb{M}$ for the typical state $z = 1$ are given by $\bv \pm \sqrt{\bv^2-4(\bv')^2-\frac{\omega^2}{4}} > 0$, so that the symmetric solution is always stable in this case. As $z$ moves away from $1$, the profile becomes unstable if
\begin{align}
\det\mathbb{M} = 4 [(\bv')^2-\bv\bv'']z^2+4\bv\bv''z+\frac{\omega^2}{4} = 0,
\end{align}
whose roots are given by
\begin{align}
z = z_\pm^* = \frac{2\bv\bv''\pm\sqrt{4\bv^2(\bv'')^2+\omega^2[\bv\bv''-(\bv')^2]}}{4\left[\bv\bv''-(\bv')^2\right]}.
\end{align}

For a DPT to occur, at least one of the roots should be real and positive. If this is the case, there are two possible scenarios:
\begin{enumerate}
\item Case of $\bv\bv''-(\bv')^2 > 0$. This case requires $\bv'' > 0$, and the only positive root is
\begin{align}
z^*_+ = \frac{2\bv\bv'' + \sqrt{4\bv^2(\bv'')^2+\omega^2[\bv\bv''-(\bv')^2]}}{4\left[\bv\bv''-(\bv')^2\right]},
\end{align}
which is always greater than $1$ and reaches the minimum at $\omega = 0$. Thus a DPT occurs due to a time-independent mode at
\begin{align} \label{eq:zc}
z = z_\rc = \frac{\bv\bv''}{\bv\bv''-\bv'^2},
\end{align}
which is always greater than $1$. Revisiting Eq.~\eqref{eq:z_def}, this implies that the symmetric (symmetry-broken) phase occupies the low-activity, low-current (high-activity, high-current) regime. A phase diagram in the $\lambda\mu$-plane corresponding to this scenario is shown in Fig.~\hyperref[fig:diagrams]{\ref*{fig:diagrams}(a)}. As will be shown later, a DPT between these two phases occurs as a second-order singularity of $\psi$ shown in Fig.~\hyperref[fig:diagrams]{\ref*{fig:diagrams}(b)}, with the optimal density $\rho_z^*$ minimizing the action exhibiting clear bifurcations shown in Fig.~\hyperref[fig:diagrams]{\ref*{fig:diagrams}(c)} and corresponding to the symmetry breaking.

\begin{figure}
\includegraphics[width=0.7\columnwidth]{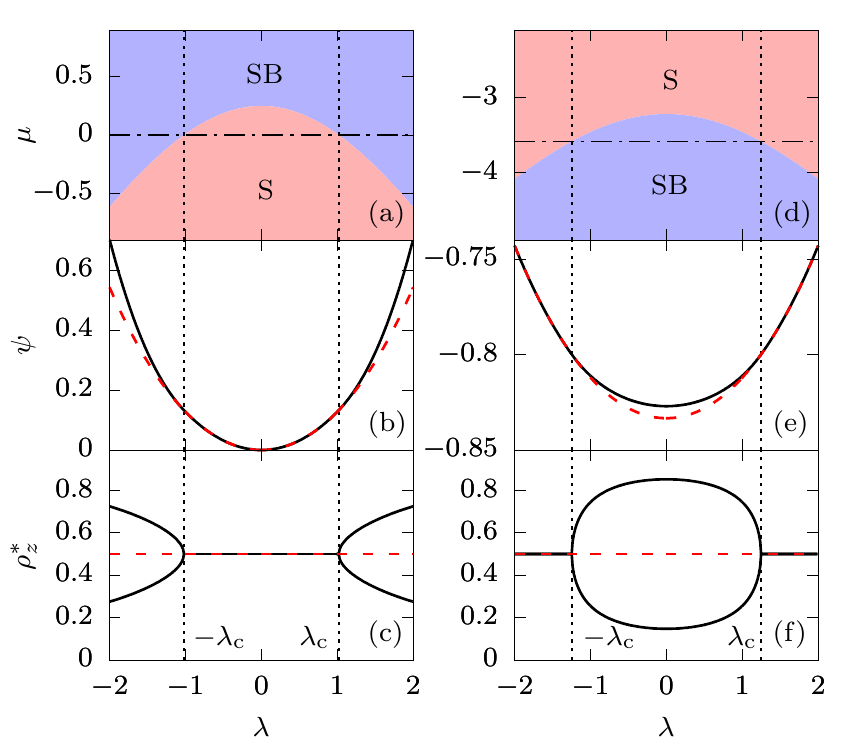}
\caption{\label{fig:diagrams} Examples of symmetry-breaking DPTs associated with current and activity large deviations of the SAP (defined in Sec.\:\ref{ssec:sap}). (a) A phase diagram for $\varepsilon = 17$ in the $\lambda\mu$-plane, with the symmetric (S) and the symmetry-broken (SB) phases indicated by different colors. (b) The scaled CGF $\psi$ of the time-averaged current for $\mu = 0$ (the dash-dotted line in (a)), which exhibits second-order singularities at $\lambda = \pm \lambda_\text{c}$ so that the actual $\psi$ (black solid line) is larger than the one corresponding to the symmetric solution (red dashed line) for $|\lambda| > \lambda_\text{c}$. (c) The optimal density of the box shows a clear symmetry breaking at $\lambda = \pm\lambda_\text{c}$. (d--f) Similar plots for $\varepsilon = -1/2$, with (e) and (f) taken along the dash-dotted line $\mu \simeq -3.58$ of (d).}	
\end{figure}

\item Case of $\bv\bv''-(\bv')^2 < 0$. Here a positive root exists if and only if $\bv'' < 0$. It is then given by
\begin{align}
z^*_- = \frac{-2\bv\bv'' + \sqrt{4\bv^2(\bv'')^2-\omega^2[(\bv')^2-\bv\bv'']}}{4\left[(\bv')^2-\bv\bv''\right]},
\end{align}
which is always less than $1$ and reaches its maximal value at $\omega = 0$. Again, a DPT occurs due to a time-independent mode at $z = z_\rc$ given by Eq.~\eqref{eq:zc}, which satisfies $0 < z_c < 1$. Combining this with Eq.~\eqref{eq:z_def}, we find that the symmetric (symmetry-broken) phase occupies the high-activity, high-current (low-activity, low-current) regime. A phase diagram in the $\lambda\mu$-plane for this scenario is illustrated in Fig.~\hyperref[fig:diagrams]{\ref*{fig:diagrams}(d)}, with second-order singularities of $\psi$ and the optimal density $\rho_z^*$ shown in Fig.~\hyperref[fig:diagrams]{\ref*{fig:diagrams}(e,f)}.
\end{enumerate}

We note that while scenario $1$ has been observed before in extended systems~\cite{HurtadoPRL2011,BaekPRL2017,BaekJPA2018}, we are not aware of any example of scenario $2$, although it bears some similarities to the DPTs of the WASEP with open boundaries~\cite{BaekPRL2017,BaekJPA2018,PerezEspigaresPRE2018} if one shifts $\lambda$ and $\mu$ appropriately. In all scenarios, a symmetry-breaking DPT occurs due to a time-independent mode. 

We next derive a Landau theory from first principles to describe the nature of the DPT in detail.

\subsection{Exact Landau theory for dynamical phase transitions}
\label{ssec:landau_theory}

Having shown that the DPTs are induced by time-independent modes, Eqs.~\eqref{eq:action_r} and \eqref{eq:psi_saddle_approx} imply that the scaled CGF takes the form
\begin{align} \label{eq:dpt_opt}
\psi(z) = \sup_{\rho,\hrho} H_z(\rho,\hrho) = \psi^\text{sym}(z) - \inf_m L_z(m),
\end{align}
where $\rho$ and $\hrho$ are time-independent solutions of Hamilton's equations~\eqref{eq:drdt} and \eqref{eq:drhdt}, and $m = \rho - 1/2$ is an order parameter quantifying the broken particle-hole symmetry. In the vicinity of a DPT, where $\epsilon_z = (z-z_\rc)/z_\rc$ is of order $m^2$, one can straightforwardly check that
\begin{align}
\rho = \frac{1}{2} + m, \quad \hrho = -\frac{\bv'}{\bv}\,m
\end{align}
yields a time-independent solution of Eqs.~\eqref{eq:drdt} and \eqref{eq:drhdt} up to order $m^2$. Using this solution in Eq.~\eqref{eq:dpt_opt} and expanding in $m$, we obtain
\begin{align} \label{eq:landau}
L_z(m) = -a\epsilon_z m^2 + b\,m^4
\end{align}
with the coefficients
\begin{align} \label{eq:landau_coeffs}
a \equiv \gamma \bv'', \quad
b \equiv \gamma\bv'z_\rc\left(\frac{1}{4}\frac{\bv'^3}{\bv^3}+\frac{1}{3}\frac{\bv^{(3)}}{\bv}-\frac{1}{2}\frac{\bv'\bv''}{\bv^2}-\frac{\bv'\bv^{(4)}}{\bv\bv''}\right).
\end{align}
The solution satisfies Eq.~\eqref{eq:dpt_opt} up to order $m^4$. This expression provides an exact Landau theory for the symmetry-breaking DPT near $z = z_\rc$ under the condition that $b > 0$ --- by tracking the optimal value of the order parameter $m = m^*_z$ minimizing $L_z$, one observes a bifurcation of $m^*_z$ and an associated jump discontinuity of $\psi''(z)$ at $z = z_\rc$ (with the locations of symmetric and symmetry-broken phases determined by the sign of $\bar v''$, as discussed above), see Fig.~\ref{fig:diagrams}. If $b<0$, one needs to expand Eq.~\eqref{eq:dpt_opt} to higher order in $m$. Note that, depending on the sign of $a$, both scenario~$1$ and scenario~$2$ described in Sec.~\ref{ssec:stability_analysis} are captured by the Landau theory.

The Landau theory obtained above has the same form as the one describing symmetry-breaking DPTs in extended systems~\cite{BaekPRL2017,BaekJPA2018}. Thus the universal features of such DPTs are captured by our large-$N$ single-box models, whose only degree of freedom plays the role of the largest-wavelength mode in extended systems. Below we explicitly construct a single-box model motivated by the Katz--Lebowitz--Spohn  (KLS) model~\cite{KatzJSP1984} which illustrates the phenomenology described so far.

Next, we examine the statistics of finite-frequency modes, which contains crucial information about the relaxation of the system near the transition. In particular, we find a behavior analogous to critical slowing down.

\subsection{Critical slowing down}
\label{ssec:crit_slow}

Let us define $\epsilon_z \equiv (z-z_\rc)/z_\rc$.
In the symmetric phase (for $\bv''\epsilon_z < 0$), from Eqs.~\eqref{eq:path_integ_r}, \eqref{eq:S_exp}, and \eqref{eq:d2S}, we find that the Gaussian fluctuations around $\rho = 1/2$ are characterized by the probability distribution
\begin{align} \label{eq:prob_phi}
P_z[\varphi] = \int \mathcal{D}\hphi \,\mathrm{e}^{-N\delta^2 S_z[\varphi,\hphi]} \sim \exp\left[-\frac{N\gamma}{8\bv z}\int\frac{d\omega}{2\pi}\,(\omega^2+\tau_z^{-2})\,\varphi_\omega\varphi_{-\omega}\right],
\end{align}
where 
\begin{align} \label{eq:tau_z}
\tau_z \equiv \sqrt{\frac{1}{16\bv z |\bv''\epsilon_z|}} \sim |\epsilon_z|^{-1/2}
\end{align}
has dimension of time. In the frequency space, the variance of the above distribution is given by
\begin{align}
\langle \varphi_\omega\varphi_{\omega'} \rangle_z = \frac{8\pi\bv z}{N\gamma}\frac{1}{\omega^2+\tau_z^{-2}}\,\delta(\omega+\omega'),
\end{align}
where $\langle\cdot\rangle_z$ denotes an average over the ensemble biased by $z$. After applying the Fourier transform, the temporal correlations are obtained as
\begin{align}
\langle \varphi(t)\varphi(t') \rangle_z = \frac{2\bv z \tau_z}{N\gamma}\,\mathrm{e}^{-|t-t'|/\tau_z} \;.
\end{align}
Thus $\tau_z$ is clearly interpreted as a correlation time, and its divergent behavior $\tau_z \sim |\epsilon_z|^{-1/2}$ near a DPT implies critical slowing down. While this derivation is valid only in the symmetric phase, it is natural to expect that the same scaling behavior will still hold in the symmetry-breaking phase.

\subsection{Example of symmetry breaking: Symmetric Antiferromagnetic Process}
\label{ssec:sap}

The KLS model is defined on a lattice where each site is occupied by at most one particle. The dynamics of the particles depend on nearest-neighbor interactions. Recently, it was shown that the KLS model, when connected to two reservoirs, exhibits a DPT when the interactions are sufficiently strongly antiferromagnetic~\cite{BaekPRL2017}. In this case, the particles prefer a profile with only every second site occupied, which amounts to having a density $\rho = 1/2$. Then the noise strength in the dynamics is found to have a local minimum at $\rho = 1/2$. To mimic this behavior, we study a single-box model with the hopping rates
\begin{align} \label{eq:SAP_rates}
W_\rR(n_1,n_2) &= n_1(N-n_2)\left[N^2+\frac{\varepsilon}{4}(n_1+n_2-N)^2\right], \nonumber\\
W_\rL(n_1,n_2) &= n_2(N-n_1)\left[N^2+\frac{\varepsilon}{4}(n_1+n_2-N)^2\right],
\end{align}
with $\varepsilon > 0$. These rates fulfill the conditions for the particle-hole symmetry and the bounded range of occupancy given in Eqs.~\eqref{eq:p-h_sym_rates} and \eqref{eq:rate_n_constraint}. They also ensure that the hopping rate attains a local minimum when the two sites involved have an average occupancy $\frac{n_1+n_2}{2} = \frac{N}{2}$. For this reason, we refer to this model as the Symmetric Antiferromagnetic Process (SAP).

For large $N$, we can use Eqs.~\eqref{eq:nu}, \eqref{eq:factorize}, and \eqref{eq:rescaling} with $k = 4$ to describe the model in terms of the rescaled parameters
\begin{align}
v(\rho) \equiv (1-\rho)\left[1+\varepsilon\left(\rho-\frac{1}{2}\right)^2\right],
\quad \alpha = \frac{1}{2}, \quad \nu = 1.
\end{align}
By Eqs.~\eqref{eq:z_def} and \eqref{eq:zc}, we obtain
\begin{align} \label{eq:z_SAP_def}
z = \rexp^{\mu/2}\cosh \frac{\lambda}{2},
\quad z_\rc = \frac{\varepsilon}{\varepsilon-2}.
\end{align}
The corresponding Landau theory is derived from Eq.~\eqref{eq:landau} as
\begin{align}
L(m) = -\varepsilon\,\epsilon_z\,m^2 + \frac{2\,\varepsilon\,(1+\varepsilon)}{\varepsilon-2}m^4.
\end{align}
Thus, if $\varepsilon > 2$ so that the coefficient of $m^4$ is positive, the model exhibits symmetry-breaking DPTs with the symmetry-broken phase occupying the high-current, high-activity regime. An example was already shown for $\varepsilon = 17$ in Fig.~\hyperref[fig:diagrams]{\ref*{fig:diagrams}(a--c)}. We again stress that this Landau theory is a direct analogue of the one describing the symmetry-breaking DPT of the KLS model in extended systems.

Interestingly, if we generalize the model to negative values of $\varepsilon$ (allowing the interactions to be {\em ferromagnetic}), the Landau theory predicts symmetry-breaking DPTs for $-1 < \varepsilon < 0$ as well. In this case, as illustrated for $\varepsilon = -1/2$ in Fig.~\hyperref[fig:diagrams]{\ref*{fig:diagrams}(d--f)}, the symmetry-broken phase corresponds to the low-current, low-activity regime. For the sake of brevity, through the rest of this paper, we shall focus on the proper SAP with $\varepsilon > 2$; however, all the results we discuss below are also easily applicable to the DPTs for $-1 < \varepsilon < 0$.

\section{Effects of finite $T$ or $N$}
\label{sec:scaling_theory}

The simplicity of the single-box model provides a convenient avenue for addressing the effects of finite $T$ or $N$ on the symmetry-breaking DPTs, which are the main subject of this section. First, taking $N\to\infty$ but leaving $T$ finite, we calculate analytically  the optimal trajectory from a given initial state and show how its final point scales with $T$ as the system approaches a symmetry-breaking DPT. Second, we consider the case $T\to\infty$ with $N$ finite and identify the exponents governing the finite-$N$ critical scalings near the DPT. These results allow us to build a comprehensive scaling theory near a symmetry-breaking DPT for finite $T$ and $N$.

\subsection{$N\to\infty$, finite $T$}
\label{ssec:fin-T}

\subsubsection{Formulation of the problem}

%In this case, the action is still dominated by smooth saddle-point trajectories. %
Near a DPT we only need to consider trajectories which are close to the symmetric solution~\eqref{eq:sym_saddle}. With these considerations in mind, it is convenient to perform a canonical change of variables
\begin{align} \label{eq:canon_coords}
\rho = \varphi + \frac{1}{2}, \quad
\hrho = \hphi -\frac{\bv'}{\bv}\varphi .
\end{align}
Since the transformation has a unit Jacobian, it does not introduce any additional term in the action. Thus, using Eqs.~\eqref{eq:action_r} and \eqref{eq:hamil_r}, the leading-order correction to the action arising from nonzero $\varphi$ and $\hphi$ is obtained as
\begin{align} \label{eq:dSz}
\Delta S_z[\varphi,\hphi] &\equiv S_z\!\left[\frac{1}{2}+\varphi,-\frac{\bv'}{\bv}\varphi+\hphi\right] - S_z\!\left[\frac{1}{2},0\right] \nonumber\\
&= -\frac{\bv'}{\bv}\frac{\varphi(T)^2-\varphi(0)^2}{2} + \tilde{S}_z[\varphi,\hphi],
\end{align}
where
\begin{align}
\tilde{S}_z[\varphi,\hphi] \equiv \int_0^{T} dt\, \left[\hphi\dot{\varphi}-h(\varphi,\hphi)\right]
\end{align}
with the effective Hamiltonian
\begin{align} \label{eq:h_def}
h(\varphi,\hphi) \equiv H_z\!\left(\frac{1}{2}+\varphi,-\frac{\bv'}{\bv}\varphi + \hphi\right) - H_z\!\left(\frac{1}{2},0\right).
\end{align}
Our goal is to minimize $\Delta S_z[\varphi,\hphi]$ for given values of $z$ and $\varphi(0)$, the value of $\varphi$ at time $t=0$. In other words, we first find the action of the optimal Hamiltonian trajectory from $\varphi(0)$ to $\varphi(T)$ with the latter allowed to take any value; then, among all such trajectories, we choose the value of $\varphi(T)$ which gives the minimal action.

\subsubsection{Exact calculation of the optimal final point}

To carry out the calculation of $\varphi(T)$, we write the variations of $\Delta S_z$ for fixed $\varphi(0)$ and $\varphi(T)$:
\begin{align}
\left.\delta\Delta S_z[\varphi,\hphi]\right|_{\varphi(T)} = \int_0^T dt\, \left\{\left(\dot\varphi-\frac{\partial h}{\partial \hphi}\right)\,\delta\hphi - \left(\dot\hphi + \frac{\partial h}{\partial\varphi}\right)\delta\varphi\right\}.
\end{align}
This gives us as expected Hamilton's equations
\begin{align} \label{eq:dphdt}
\dot\varphi = \frac{\partial h}{\partial \hphi}, \quad \dot\hphi = -\frac{\partial h}{\partial \varphi}.
\end{align}
Then, using Eq.~\eqref{eq:dSz} and allowing variations of $\varphi(T)$, we obtain
\begin{align}
\delta\Delta S_z[\varphi,\hphi] = \left[\hphi(T)-\frac{\bv'}{\bv}\varphi(T)\right]\delta\varphi(T) + \left.\delta\Delta S_z[\varphi,\hphi]\right|_{\varphi(T)}.
\end{align}
This implies that, among all the solutions of Eq.~\eqref{eq:dphdt}, the one with the minimal action satisfies
\begin{align} \label{eq:phi_opt_cond}
\hphi(T) = \frac{\bv'}{\bv}\varphi(T).
\end{align}
To proceed, we note that the above relation gives a conserved ``mechanical energy'' of the Hamiltonian dynamics as a function of $\varphi(T)$:
\begin{align} \label{eq:mech_energy}
E(\varphi(T)) \equiv h\!\left(\varphi(T),\frac{\bv'}{\bv}\varphi(T)\right).
\end{align}
With this the minimum of $\Delta S_z$ can be written as
\begin{align}
\inf_{\varphi,\hphi} \Delta S_z[\varphi,\hphi] = \inf_{\varphi(T)} \left[-\frac{\bv'}{\bv}\frac{\varphi(T)^2-\varphi(0)^2}{2} + \int_{\varphi(0)}^{\varphi(T)}\!\!\! d\varphi \,\hphi \ - \ E(\varphi(T))T\right].
\end{align}
Differentiating the rhs with respect to $\varphi(T)$ and using Eq.~\eqref{eq:phi_opt_cond}, we find that the minimal $\Delta S_z$ requires
\begin{align} \label{eq:phiT_opt_cond_0}
\int_{\varphi(0)}^{\varphi(T)} d\varphi \,\frac{\partial\hphi}{\partial\varphi(T)} - E'(\varphi(T))\,T = 0.
\end{align}
In the following discussions, the optimal $\varphi(T)$ is obtained by solving this equation.

\subsubsection{Numerical results for the SAP}

\begin{figure}
\includegraphics[width=0.7\textwidth]{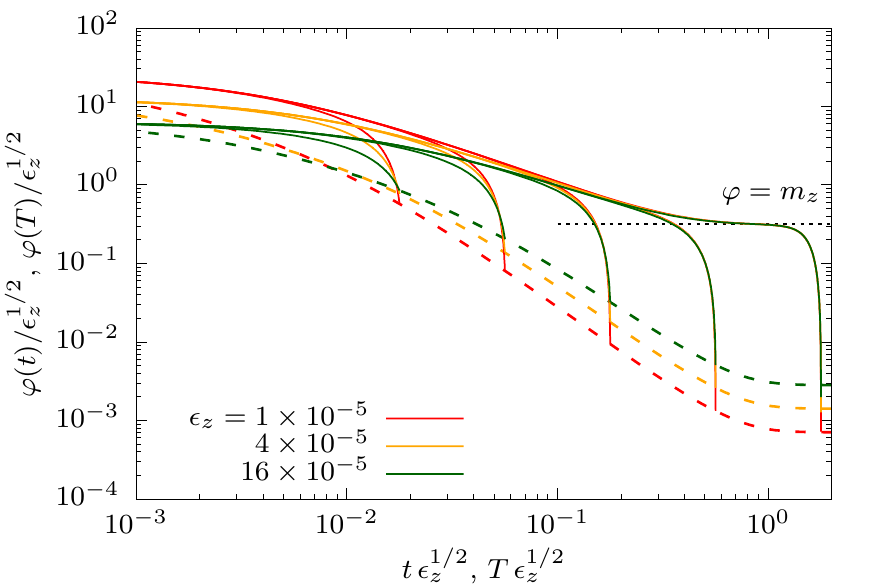}
\caption{\label{fig:phit_t} Infinite-$N$, finite-$T$ relaxation trajectories of the SAP with $\varepsilon = 4$ near a DPT. Solid curves: saddle-point trajectories from the initial state $\varphi(0) = 0.08$ and varied values of $T$. Dashed curves: final state $\varphi(T)$ reached by the saddle-point trajectories. Both types of curves share the same color scheme.}	
\end{figure}

\begin{figure}
\includegraphics[width=\textwidth]{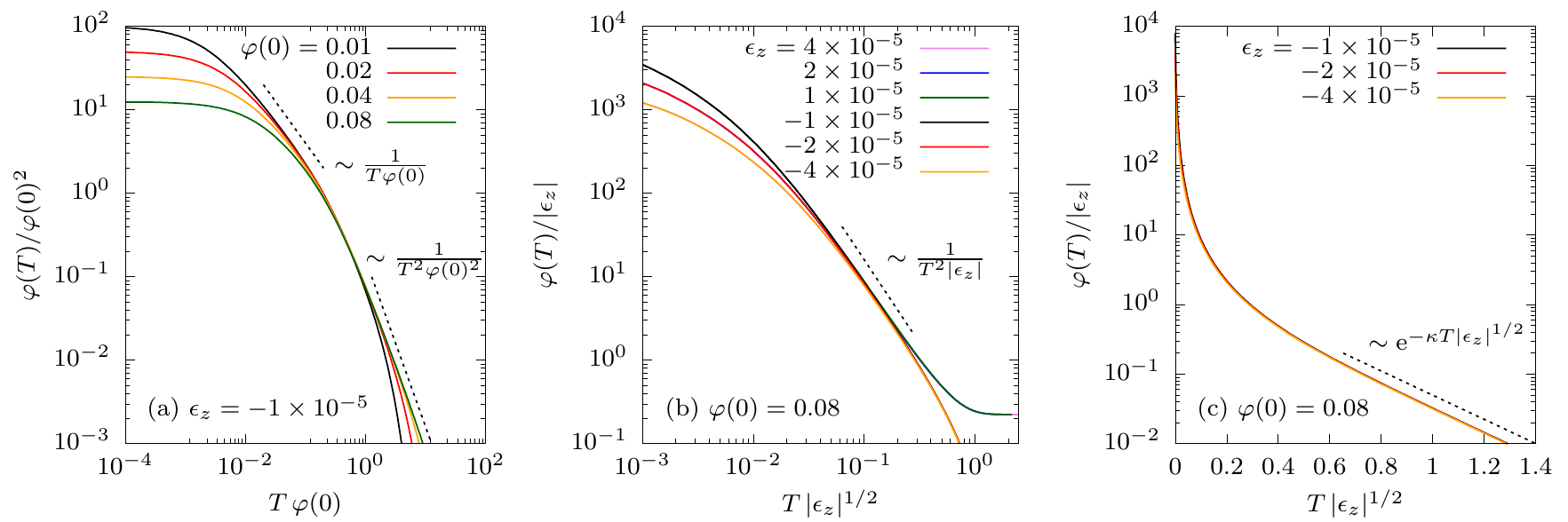}
\caption{\label{fig:phiT_T_scaling} Finite-$T$ scaling behaviors of the final state $\varphi(T)$ reached by the SAP with $\varepsilon = 4$. (a) If $T$ is small, $\varphi(T)$ is governed by the initial state $\varphi(0)$. (b) For intermediate values of $T$, $\varphi(T)$ shows a power-law decay governed by $|\epsilon_z|$, irrespective of the sign $\epsilon_z$. (c) In the symmetric phase, $\varphi(T)$ exhibits an exponential decay if $T$ is large enough.}	
\end{figure}

With Eqs.~\eqref{eq:drdt}, \eqref{eq:drhdt}, and \eqref{eq:phiT_opt_cond_0}, we are ready to calculate the optimal finite-$T$ trajectories for given $z$ and $\varphi(0)$. We first consider numerical solutions and identify different scaling regimes, each of which will be described by analytical arguments later. In Fig.~\ref{fig:phit_t}, we illustrate such trajectories for the SAP with $\varepsilon = 4$ in the symmetry-broken phase, all of them starting from the initial state $\varphi(0) = 0.08$ while the values of $z$ and $T$ are varied. The optimal trajectories themselves are marked by solid curves, whereas their final-time value~$\varphi(T)$ is shown as a dashed curves as $T$ changes continuously. Notably, if $T$ is sufficiently large, the trajectories initially appear to saturate at the value of the order parameter $m_z \sim \epsilon_z^{1/2}$; however, they eventually move past the plateau (with a characteristic time scale which, as shown below, reflects the critical slowing down $\tau_z \sim \epsilon_z^{-1/2}$) and end up much closer to the symmetric state $\rho = 1/2$. As is evident from the data collapse, $\varphi(t)$ and $\varphi(T)$ exhibit different scaling behaviors near a DPT.

In Fig.~\ref{fig:phiT_T_scaling}, using the SAP with $\varepsilon = 4$, we show that $\varphi(T)$ exhibits three different scaling regimes depending on the duration of the observation period $T$:
\begin{itemize}
\item {\it Regime I}. If the observation period is not long enough, the initial state $\varphi(0)$ heavily influences the entire trajectory, including the final state $\varphi(T)$ obeying
\begin{align}
\varphi(T) &\sim \varphi(0)/T \quad \text{for $T \ll \varphi(0)^{-1}$.} \label{eq:regime_1}
\end{align}
The above scaling behavior is shown in Fig.~\hyperref[fig:phiT_T_scaling]{\ref*{fig:phiT_T_scaling}(a)}.
\item {\it Regime II}. As the observation period becomes longer, the initial-state dependence starts to disappear after a time scale $\varphi(0)^{-1}$, beyond which proximity to the critical point becomes manifest in the power-law decay
\begin{align}
\varphi(T) &\sim T^{-2} \quad \text{for $\varphi(0)^{-1} \ll T \ll |\epsilon_z|^{-1/2}$,} \label{eq:regime_2}
\end{align}
as also shown in the middle section of Fig.~\hyperref[fig:phiT_T_scaling]{\ref*{fig:phiT_T_scaling}(b)}. At this stage, there is no distinction between the symmetric ($\epsilon_z < 0$) and symmetry-broken ($\epsilon_z > 0$) phases.
\item {\it Regime III}. When $T$ is sufficiently larger than the correlation time scale $\tau_z \sim |\epsilon_z|^{-1/2}$, $\varphi(T)$ converges exponentially to zero in the symmetric phase (see Fig.~\hyperref[fig:phiT_T_scaling]{\ref*{fig:phiT_T_scaling}(c)}) and to nonzero values in the symmetry-broken phase (see Fig.~\hyperref[fig:phiT_T_scaling]{\ref*{fig:phiT_T_scaling}(b)}), as we show below:
\begin{align}
\varphi(T) &\sim |\epsilon_z|\, e^{-2\sqrt{\gamma\bv z_\rc|a\epsilon_z|}T}& &\text{for $T \gg |\epsilon_z|^{-1/2}$ and $a\epsilon_z < 0$,}\nonumber\\
\lim_{T\to\infty} \varphi(T) &\simeq \frac{a\epsilon_z}{2\sqrt{bc}}& &\text{for $T \gg |\epsilon_z|^{-1/2}$ and $a\epsilon_z > 0$.}\label{eq:regime_3}	
\end{align}
\end{itemize}

Based on these scaling behaviors, one can infer the following scaling forms describing the crossovers between adjacent scaling regimes:
\begin{align} \label{eq:phi_FSS}
	\varphi(T) &= \begin{cases}
\varphi(0)^{2} \,\mathcal{F}_1(T\,\varphi(0)) \quad \text{between regimes I and II}, \\ 	
|\epsilon_z| \,\mathcal{F}_2(T^2\,a\epsilon_z) \quad \text{between regimes II and III}.
 \end{cases}
\end{align}
To be consistent with the scaling behaviors in each regime, the functions $\mathcal{F}_1$ and $\mathcal{F}_2$ should satisfy 
\begin{align}
\mathcal{F}_1(x) &\sim \begin{cases}
 1/x &\text{ for $|x| \ll 1$,}\\
 1/x^2 &\text{ for $|x| \gg 1$,}
 \end{cases} \\
\mathcal{F}_2(x) &\sim \begin{cases}
 1/x &\text{ for $|z| \ll 1$,}\\
 e^{-\text{const.}\times \sqrt{|x|}} &\text{ for $|x| \gg 1$ and $x < 0$,}\\
 \text{const.} &\text{ for $x \gg 1$ and $x > 0$.}
 \end{cases}
\end{align}
The existence of such $\mathcal{F}_1$ ($\mathcal{F}_2$) is manifest in the data collapse(s) shown in Fig.~\hyperref[fig:phiT_T_scaling]{\ref*{fig:phiT_T_scaling}(a)} (Fig.~\hyperref[fig:phiT_T_scaling]{\ref*{fig:phiT_T_scaling}(b, c)}).

Due to the simplicity of the single-box models, all the numerical results discussed above can be theoretically derived from first principles, as we now show.

\subsubsection{Derivation of the scaling theory}

\begin{figure}
\includegraphics[width=0.7\columnwidth]{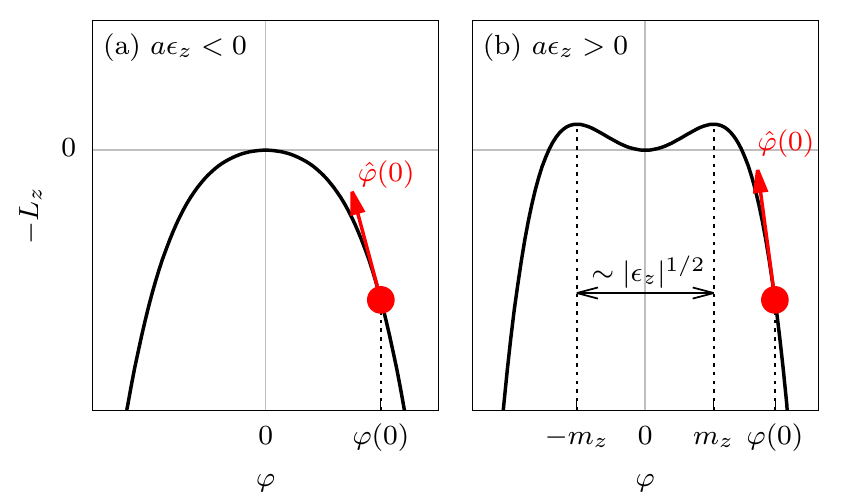}
\caption{\label{fig:potential} The unstable quartic potential governing the infinite-$N$, finite-$T$ saddle-point trajectories in (a) the symmetric and (b) the symmetry-broken phases.}
\end{figure}

To analytically calculate $\varphi(T)$ satisfying Eq.~\eqref{eq:phiT_opt_cond_0}, one needs to examine the form of the Hamiltonian $h(\varphi,\hphi)$. In what follows, we approximate $h(\varphi,\hphi)$ by using Eq.~\eqref{eq:hamil_r} in Eq.~\eqref{eq:h_def} and expanding the latter for small $\varphi$ and $\hphi$ to obtain
\begin{align} \label{eq:h_approx}
h(\varphi,\hphi) = c\hphi^2-L_z(\varphi) +o\!\left(\hphi^2,\epsilon_z\varphi^2,\varphi^4\right),
\end{align}
where $c \equiv \gamma\bv z_\rc$ and $L_z$ are as defined in Eqs.~\eqref{eq:landau} and \eqref{eq:landau_coeffs}, respectively.
As we show, the results below are unaffected by the neglected higher-order terms. This approximate formula has a convenient interpretation as the Hamiltonian of a Newtonian particle of mass $\frac{1}{2c}$, velocity $\hphi$ and position $\varphi$ in an unstable quartic potential $-L_z(\varphi)$, represented schematically in Fig.~\ref{fig:potential}.

Using Eqs.~\eqref{eq:h_def} and \eqref{eq:mech_energy}, the energy conservation $h(\varphi,\hphi) = E(\varphi(T))$ implies
\begin{align}
\hphi = \hphi_\pm \equiv \pm\sqrt{\frac{E(\varphi(T)) + L_z(\varphi)}{\gamma\bv z_\rc}}.
\end{align}
Near a symmetry-breaking DPT, it is natural to expect that the optimal trajectory stays close to the symmetric solution~\eqref{eq:sym_saddle}. Thus the initial velocity should be in the uphill direction. For generic situations near the DPT, we expect $\varphi(0)$ to be well within the unstable branches of the potential (i.e., $|\varphi(0)| \gg |\epsilon_z|^{1/2}$), see Fig.~\ref{fig:potential}. In this case, the sign of $\hphi(0)$ should be opposite to that of $\varphi(0)$. Since the system satisfies a particle-hole symmetry, without loss of generality, we can focus on the case where $\varphi(0) > 0$, so that $\hphi(0) < 0$ and $\hphi = \hphi_-$. 

Using the above relation and Eq.~\eqref{eq:phiT_opt_cond_0}, we obtain
\begin{align}
2\sqrt{\gamma\bv z_\rc}\,T \simeq \int_{\varphi(T)}^{\varphi(0)} d\varphi \, \frac{1}{\sqrt{E(\varphi(T)) + L_z(\varphi)}}.
\end{align}
This can be further simplified to 
\begin{align} \label{eq:T_integ}
2\sqrt{\gamma\bv z_\rc}\,T \simeq \int_{\varphi(T)}^{\varphi(0)} d\varphi \,\frac{1}{\sqrt{c\,\varphi(T)^2 - a\epsilon_z\varphi^2 + b\,\varphi^4}}
\end{align}
by using Eq.~\eqref{eq:landau} and noting that Eqs.~\eqref{eq:mech_energy} and \eqref{eq:h_approx} give
\begin{align}
E(\varphi(T)) \simeq \frac{\gamma \bv'^2 z_\rc}{\bv}\varphi(T)^2	 - L_z(\varphi(T)) \simeq \frac{\gamma \bv'^2 z_\rc}{\bv}\varphi(T)^2,
\end{align}
where the second approximation is due to the quartic potential $L_z(\varphi(T)) = O\!\left(\epsilon_z\varphi(T)^2,\varphi(T)^4\right)$ being negligible compared to the ``kinetic'' component near the DPT where $\epsilon_z \ll 1$. Depending on which term in the denominator dominates the integral in Eq.~\eqref{eq:T_integ}, we identify the following three scaling regimes in order of increasing $T$:

{\em Regime I.} --- Suppose that the integral in Eq.~\eqref{eq:T_integ} is dominated by contributions from $c\,\varphi(T)^2$. Then, using a Taylor expansion, Eq.~\eqref{eq:T_integ} can be approximated as
\begin{align}
T = \frac{\varphi(0)}{2\gamma\bv'z_\rc\varphi(T)} + O\!\left(\frac{\epsilon_z\varphi(0)^3}{\varphi(T)^3},\frac{\varphi(0)^5}{\varphi(T)^3}\right)
\end{align}
implying $\varphi(T) \sim \varphi(0)/T$. This scaling behavior is self-consistent if and only if the latter two terms on the rhs are much smaller than $T$, which requires $T \ll |\epsilon_z|^{-1/2}$ and $T \ll 1/\varphi(0)$. Since we have already assumed $\varphi(0) \gg |\epsilon_z|^{1/2}$, the latter condition is automatically implied by the former. Therefore
\begin{align}
\varphi(T) \sim \frac{\varphi(0)}{T} \quad\text{for $T \ll \frac{1}{\varphi(0)}$,}
\end{align}
which is the same as Eq.~\eqref{eq:regime_1}.

{\em Regime II.} --- Suppose that the integral in Eq.~\eqref{eq:T_integ} is dominated by contributions from $b\,\varphi^4$, which requires $\varphi \gg \max[\sqrt{\frac{a\epsilon_z}{b}},\,\left(\frac{c}{b}\right)^{1/4}\sqrt{\varphi(T)}]$. Thus Eq.~\eqref{eq:T_integ} can be approximated as
\begin{align} \label{eq:regime_2_integral}
2\sqrt{\gamma\bv z_\rc}\,T \simeq \frac{1}{\sqrt{b}}\int_{\max[\sqrt{\frac{a\epsilon_z}{b}},\,\left(\frac{c}{b}\right)^{1/4}\sqrt{\varphi(T)}]}^{\varphi(0)}\frac{1}{\varphi^2}
\simeq \frac{1}{\sqrt{b}\max[\sqrt{\frac{a\epsilon_z}{b}},\,\left(\frac{c}{b}\right)^{1/4}\sqrt{\varphi(T)}]} - \frac{1}{\sqrt{b}\,\varphi(0)}.
\end{align}
For the moment, we assume that the dominating term on the rhs is given by the second argument of $\max[\cdot]$, so that
\begin{align}
2\sqrt{\gamma\bv z_\rc}\,T \simeq \frac{1}{(cb)^{1/4}\sqrt{\varphi(T)}},
\end{align}
which yields $\varphi(T) \sim T^{-2}$. This is self-consistent if $T \ll |\epsilon_z|^{-1/2}$ (by comparison between $\sqrt{\frac{a\epsilon_z}{b}}$ and $\left(\frac{c}{b}\right)^{1/4}\sqrt{\varphi(T)})$ in the $\max[\cdot]$) and if $T \gg \frac{1}{\varphi(0)}$ (so that $
\frac{1}{\sqrt{b}\,\varphi(0)}$ can be neglected). Therefore we obtain a scaling regime
\begin{align}
\varphi(T) \sim \frac{1}{T^2} \quad\text{for $\frac{1}{\varphi(0)} \ll T \ll \frac{1}{\sqrt{|\epsilon_z|}}$,}
\end{align}
which is identical to \eqref{eq:regime_2}. It is straightforward to show that other choices of dominating terms in Eq.~\eqref{eq:regime_2_integral} do not lead to self-consistent results.

{\em Regime III.} --- Finally, we consider the case where the contribution from $\epsilon_z \varphi^2$ is not negligible. Depending on the sign of $a\epsilon_z$, it is natural to divide this regime into two different cases. For $a\epsilon_z < 0$ (inside the symmetric phase), the integral in Eq.~\eqref{eq:T_integ} can be dominated solely by $\epsilon_z \varphi^2$. Since $\epsilon_z\varphi^2 \gg c\,\varphi(T)^2 + b\,\varphi^4$ requires the range of the integral to satisfy $\sqrt{\frac{c}{|a\epsilon_z|}}\varphi(T) \ll \varphi \ll \sqrt{\frac{|a\epsilon_z|}{b}}$, Eq.~\eqref{eq:T_integ} can be approximated as
\begin{align}
	2\sqrt{\gamma\bv z_\rc}\,T \simeq \int_{\sqrt{\frac{c}{|a\epsilon_z|}}\varphi(T)}^{\sqrt{\frac{|a\epsilon_z|}{b}}} d\varphi\, \frac{1}{\sqrt{|a\epsilon_z|}\,\varphi} \simeq \frac{1}{\sqrt{|a\epsilon_z|}} \ln \frac{|a\epsilon_z|}{\sqrt{bc} \,\varphi(T)},
\end{align}
implying $\varphi(T) \sim |\epsilon_z|\, e^{-2\sqrt{\gamma\bv z_\rc|a\epsilon_z|}T}$. This scaling behavior is consistent with the range of the above integral if and only if $T \gg |\epsilon_z|^{-1/2}$. Therefore
\begin{align}
\varphi(T) \sim |\epsilon_z|\, e^{-2\sqrt{\gamma\bv z_\rc|a\epsilon_z|}T} \quad \text{for $a\epsilon_z < 0$ and $T \gg |\epsilon_z|^{-1/2}$},
\end{align}
which reproduces the first part of Eq.~\eqref{eq:regime_3}. On the other hand, if $a\epsilon_z > 0$, we have
\begin{align}
\inf_\varphi L_z(\varphi) = -\frac{a^2\epsilon_z^2}{4b} < 0.
\end{align}
For Eq.~\eqref{eq:T_integ} to be consistent with positive and arbitrarily large $T$, the value of $\varphi(T)$ must be such that the denominator of the integrand in Eq.~\eqref{eq:T_integ} remains positive but approaches arbitrarily close to zero in some part of the trajectory. Thus, $\varphi(T)$ eventually converges to a nonzero value
\begin{align} \label{eq:subcrit_limit}
\lim_{T\to\infty} \varphi(T) \simeq \frac{a\epsilon_z}{2\sqrt{bc}} \quad \text{for $a\epsilon_z > 0$},
\end{align}
which is in agreement with the second part of Eq.~\eqref{eq:regime_3}. As was already shown in Fig.~\ref{fig:phit_t}, this limiting value of $\varphi(T)$ is not equal to a minimum of $L_z$ located at $m_z = \sqrt{\frac{a\epsilon_z}{2b}}$ but satisfies $0 < \varphi(T) < m_z < \varphi(0)$. Even then, the integral in Eq.~\eqref{eq:T_integ} is dominated by the interval satisfying $\varphi \simeq m_z$, where the denominator of the integrand is very small. This implies that, as $T$ becomes larger, the trajectory stays close to $m_z$ for a longer period of time, as clearly shown in Fig.~\ref{fig:phit_t}.

These derivations fully justify the scaling behaviors stated in Eqs.~\eqref{eq:regime_1}, \eqref{eq:regime_2}, and \eqref{eq:regime_3}. Since the Landau-theory approach we have followed is rather general, we expect that similar behaviors will be observed not only in the DPTs of the single-box SAP, but in the broader range of the generic symmetry-breaking DPTs described in Sec.~\ref{sec:symbreak}.

\subsection{$T\to\infty$, finite $N$}
\label{ssec:fin-N}

\subsubsection{General formalism}

In this case, one cannot rely on the saddle-point method as fluctuations are not negligible. Instead, we consider the limit $T\to\infty$ by studying the spectral properties of the stochastic process. To this aim, we consider a vector in the Hilbert space representing a biased distribution
\begin{align}
\left|G_{\lambda,\mu}(t)\right\rangle \equiv \sum_{n=0}^N \left\langle\mathrm{e}^{t(\lambda J_t + \mu K_t)}\right\rangle_{n(t) = n}\,|n\rangle,
\end{align}
where $J_t$ and $K_t$ are as defined in Eqs.~\eqref{eq:j_def} and \eqref{eq:k_def}, respectively, and $\langle\cdot\rangle_{n(t) = n}$ denotes an average over all histories under the constraint that the box has $n$ particles at time $t$. Then it is known~(see, for example, \cite{ChetriteAHP2015}) that $\left|G_{\lambda,\mu}(t)\right\rangle$ evolves according to 
\begin{align}
\partial_t\left|G_{\lambda,\mu}(t)\right\rangle = \mathbb{W}_{\lambda,\mu}\left|G_{\lambda,\mu}(t)\right\rangle,
\end{align}
where the {\em tilted generator} $\mathbb{W}_{\lambda,\mu}$ is an ($N+1$)-by-($N+1$) matrix defined as
\begin{align} \label{eq:W_def}
\left(\mathbb{W}_{\lambda,\mu}\right)_{0,n} &\equiv
\left[\mathrm{e}^{(\mu+\lambda)/2}W_\text{R}(1,\bn_b)+\mathrm{e}^{(\mu-\lambda)/2}W_\text{L}(\bn_a,1)\right]\delta_{1,n} - \left[W_\text{R}(\bn_a,0)+W_\text{L}(0,\bn_b)\right]\delta_{0,n},\nonumber\\
\left(\mathbb{W}_{\lambda,\mu}\right)_{N,n} &\equiv
\left[\mathrm{e}^{(\mu+\lambda)/2}W_\text{R}(\bn_a,N-1)+\mathrm{e}^{(\mu-\lambda)/2}W_\text{L}(N-1,\bn_b)\right]\delta_{N-1,n}\nonumber\\
&\quad - \left[W_\text{R}(N,\bn_b)+W_\text{L}(\bn_a,N)\right]\delta_{N,n},\nonumber\\
\left(\mathbb{W}_{\lambda,\mu}\right)_{m,n} &\equiv
\left[\mathrm{e}^{(\mu+\lambda)/2}W_\text{R}(n,\bn_b)+\mathrm{e}^{(\mu-\lambda)/2}W_\text{L}(\bn_a,n)\right]\delta_{m+1,n} \nonumber\\
&\quad + \left[\mathrm{e}^{(\mu+\lambda)/2}W_\text{R}(\bn_a,n)+\mathrm{e}^{(\mu-\lambda)/2}W_\text{L}(n,\bn_b)\right]\delta_{m-1,n}\nonumber\\
&\quad - \left[W_\text{R}(n,\bn_b)+W_\text{L}(\bn_a,n)+W_\text{R}(\bn_a,n)+W_\text{L}(n,\bn_b)\right]\delta_{m,n},
\end{align}
with integer indices $m \in [1,N-1]$ and $n \in [0,N]$ and where $\delta_{i,j}$ denotes the Kronecker delta.

Let us denote by $\Lambda_0(\lambda,\mu)$ and $\Lambda_1(\lambda,\mu)$ the two eigenvalues of $\mathbb{W}_{\lambda,\mu}$ with the largest and the second largest real part, respectively. By the Perron--Frobenius theorem, $\Lambda_0(\lambda,\mu)$ is always guaranteed to be real-valued. Thus, using Eq.~\eqref{eq:Psi_def}, the scaled CGF before the rescaling by Eq.~\eqref{eq:rescaling} satisfies
\begin{align} \label{eq:Psi_L0}
\Psi(\lambda,\mu) = \Lambda_0(\lambda,\mu).
\end{align}
The Perron--Frobenius theorem also implies that the leading eigenvalue $\Lambda_0(\lambda,\mu)$ is always unique, so that $\Psi$ cannot have singularities at finite $N$. However, by examining how $\Psi$ develops a second-order singularity in $\psi$ as $N\to\infty$, one can identify the scaling exponent governing finite-$N$ effects in the $\lambda\mu$-plane. Moreover, the spectral gap
\begin{align} \label{eq:spectral_gap}
\Delta \Lambda(\lambda,\mu) \equiv \mathrm{Re}\left[\Lambda_0(\lambda,\mu)-\Lambda_1(\lambda,\mu)\right],
\end{align}
whose inverse characterizes the relaxation time scale, is also useful as it reflects the effects of finite $N$ on the critical slowing down.

\subsubsection{Exact numerical diagonalization of the SAP}
\label{sec:exactdiagSAP}

\begin{figure}
\includegraphics[width=0.7\columnwidth]{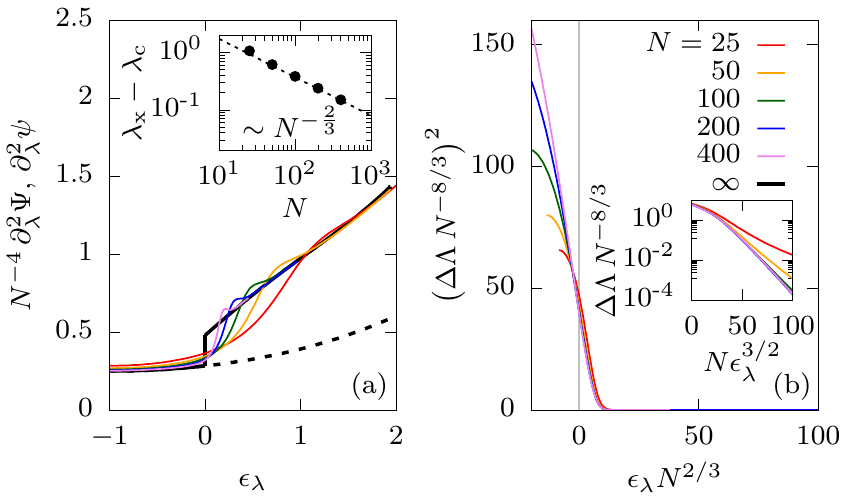}
\caption{\label{fig:psisecond} Finite-$N$ scaling behaviors obtained from the exact diagonalization of the SAP with $\varepsilon = 17$. (a) The asymptotic ($N\to\infty$) scaled CGF $\psi$ (thick black line) has a jump discontinuity in the second-order derivative when the optimal profile switches from the symmetric solution (continued by the dashed black line) to the symmetry-broken solutions. The finite-$N$ scaled CGF $\Psi$ is always smooth (thin colored lines), but approaches $\psi$ as $N$ increases. Inset: the crossing $\lambda_\mathrm{x}$ between a finite-$N$ curve and the asymptotic result approaches the DPT $\lambda_\text{c}$ according to a power law $\lambda_\mathrm{x}-\lambda_\text{c} \sim N^{-2/3}$. (b) The spectral gap $\Delta\Lambda$ is well collapsed by the finite-$N$ scaling hypothesis~\eqref{eq:DL_fss} with the scaling function satisfying Eq.~\eqref{eq:DL_fss_func}. Inset: the $\epsilon_\lambda > 0$ regime is shown in detail, revealing the exponential decrease of the gap. All plots use the same color scheme for different values of $N$.}
\end{figure}

Using the SAP hopping rates~\eqref{eq:SAP_rates} and the reservoir densities $\bn_a = \bn_b = 1/2$ in Eq.~\eqref{eq:W_def}, the tilted generator of the SAP is obtained as
\begin{align} \label{eq:W_SAP_def}
\left(\mathbb{W}_z\right)_{0,n} &\equiv N\left[N^2+\frac{\varepsilon}{4}\left(n-\frac{N}{2}\right)^2\right]\left(z\delta_{1,n}-N\delta_{0,n}\right),\nonumber\\
\left(\mathbb{W}_z\right)_{N,n} &\equiv N\left[N^2+\frac{\varepsilon}{4}\left(n-\frac{N}{2}\right)^2\right]\left(z\delta_{N-1,n}-N\delta_{N,n}\right),\nonumber\\
\left(\mathbb{W}_z\right)_{m,n} &\equiv
N\left[N^2+\frac{\varepsilon}{4}\left(n-\frac{N}{2}\right)^2\right]\left\{z\left[n\delta_{m+1,n}+(N-n)\delta_{m-1,n}\right]-N\delta_{m,n}\right\}
\end{align}
for the integer indices $m \in [1,N-1]$ and $n \in [0,N]$, where $z = z(\lambda,\mu)$ is defined as in Eq.~\eqref{eq:z_SAP_def}. In Fig.~\ref{fig:psisecond}, we show numerical results obtained from the exact diagonalization of $\mathbb{W}_z$ with $\varepsilon = 17$, which provide concrete examples of finite-$N$ effects. While all the results are restricted to the $\lambda$-axis ($\mu = 0$), it is straightforward to generalize them to the entire $\lambda\mu$-plane, as Eq.~\eqref{eq:z_SAP_def} implies that $\epsilon_\lambda$ can always be replaced with $\epsilon_z$.

In Fig.~\hyperref[fig:psisecond]{\ref*{fig:psisecond}(a)}, we show the second-order derivative of the scaled CGF $\Psi$, which is calculated from the leading eigenvalue $\Lambda_0$ by Eq.~\eqref{eq:Psi_L0}. In the $N\to\infty$ limit, as discussed below in Sec.~\ref{ssec:landau_theory}, the second derivative of asymptotic scaled CGF $\psi$ (thick black curve) has a jump discontinuity at $\epsilon_\lambda = 0$ as the symmetry is broken (for comparison, the continuation of the contribution from the symmetric solution is shown by a dashed black curve). While $\Psi$ at finite $N$ (thin colored lines) is always smooth, $N^{-4}\partial_\lambda^2\Psi$ clearly approaches $\partial_\lambda^2\psi$ as $N$ becomes larger. The inset shows that $\lambda_\mathrm{x}(N)$, defined as the value of $\lambda$ where the finite-$N$ and the asymptotic curves cross each other, converges to the DPT $\lambda = \lambda_\text{c}$ according to a power-law decay $N^{-2/3}$. We thus observe that the scale of $\epsilon_\lambda$ characterizing the onset of finite-$N$ effects is given by $\epsilon_\lambda \sim N^{-2/3}$.

In Fig.~\hyperref[fig:psisecond]{\ref*{fig:psisecond}(b)}, we show how the spectral gap of $\mathbb{W}_z$ obtained at different values of $N$ can be collapsed. As $N$ increases, one observes a collapse to a linear behavior both in the main plot and the (log-linear) inset, which implies a scaling form (after replacing $\epsilon_\lambda$ with $a\epsilon_z$)
\begin{align} \label{eq:DL_fss_SAP}
\Delta\Lambda(\epsilon_z,N) = N^{8/3}\mathcal{G}(a\epsilon_z N^{2/3}),
\end{align}
where the function $\mathcal{G}$ shows the asymptotic behaviors
\begin{align} \label{eq:DL_fss_func_SAP}
\mathcal{G}(x) \sim \begin{cases}
 	\mathrm{e}^{-c' x^{3/2} + o(x^{3/2})} &\text{ for $|x| \gg 1$ and $x > 0$},\\
 	|x|^{1/2} &\text{ for $|x| \gg 1$ and $x < 0$}
 \end{cases}
\end{align}
with a positive constant $c'$. Applying the rescaling scheme~\eqref{eq:rescaling} with $k = 4$ (as discussed in Sec.~\ref{ssec:sap}), we find the relaxation time scale at a DPT ($\epsilon_z = 0$)
\begin{align}
\tau_{z_\text{c}} \sim \frac{N^3}{\Delta\Lambda(0,N)} \sim N^{1/3}.
\end{align}
As expected the critical slowing down ({\em i.e.}, divergence of $\tau_z$ as $\epsilon_z\to 0$) is constrained by the finite value of $N$.

While these observations are based on the numerical diagonalization of the SAP, we argue that they are relevant to a broad range of symmetry-breaking DPTs induced by the same mechanism, as supported by a heuristic argument described below.

\subsubsection{Argument for finite-$N$ scaling}

To understand the finite-$N$ scaling exponents identified above, we study how the finite-$N$ corrections can become large enough to erase the second-order singularity of the scaled CGF $\psi$. Integrating the Gaussian fluctuations described by Eq.~\eqref{eq:prob_phi}, the correction to $\psi$ in the symmetric phase (corresponding to $a\epsilon_z < 0$, as explained in Sec.~\ref{ssec:stability_analysis}) is 
 \begin{align}
\delta\psi(z) = \frac{1}{NT}\ln \int \mathcal{D}\varphi \, P_z[\varphi] = -\frac{1}{NT}\int d\omega\, \ln \left[\omega^2 + \tau_z^{-2}\right].
\end{align}
As a corollary, the correction to $\psi''$ is given by
\begin{align}
\delta\psi''(z) = \frac{32\bv\bv''}{NTz_\rc}\int d\omega\, \frac{\omega^2+8\bv\bv''z_\rc-\tau_z^{-2}}{\left(\omega^2+\tau_z^{-2}\right)^2}.
\end{align}
We note that the factor $T$ in the denominator is always cancelled by the IR cutoff of the integral. Moreover, due to critical slowing down ({\em i.e.}, small $\tau_z^{-2}$ in the denominator), near a DPT the low-frequency range dominates the integral. Thus we can write
\begin{align}
\delta\psi''(z) \sim  \frac{\tau_z^3}{N} \sim \frac{|\epsilon_z|^{-3/2}}{N},
\end{align}
which implies that $\delta\psi''$ can remove the jump discontinuity of $\psi''$ only if $|\epsilon_z| \lesssim N^{-2/3}$. Assuming the scaling behavior to be homogeneous within the regime, this gives a heuristic explanation for why the finite-$N$ scaled CGF $\Psi$ converges to the asymptotic $\psi$ according to a power-law decay $N^{-2/3}$, as shown in the inset of Fig.~\hyperref[fig:psisecond]{\ref*{fig:psisecond}(a)}. We note that this argument is fully analogous to that for the finite-size scaling theory for symmetry-breaking DPTs in extended systems~\cite{BaekJPA2018}, with $N$ playing the role of the linear system size; hence the same exponent $2/3$ governs the finite-size scaling in both types of systems.

We now turn to the scaling behavior of the spectral gap $\Delta\Lambda$, whose inverse captures the dominant time scale. Close to a DPT on the side of the symmetric phase ($a\epsilon_z < 0$), if the finite-$N$ effects are negligible ($|\epsilon_z| \gg N^{-2/3}$), $\Delta\Lambda$ satisfies
\begin{align} \label{eq:DL_scaling_S}
	\Delta\Lambda(\epsilon_z,N) \sim N^{k-1}|\epsilon_z|^{1/2}.
\end{align}
Here $N^{k-1}$ stems from the rescaling of time shown in Eq.~\eqref{eq:rescaling}, and $|\epsilon_z|^{1/2}$ reflects the critical slowing down $\tau_z \sim |\epsilon_z|^{-1/2}$. On the other hand, if we approach a DPT from the side of the symmetry-broken phase ($a\epsilon_z > 0$) while keeping outside the finite-$N$ scaling regime, the intermittent flipping between the two symmetry-broken solutions $\varphi = \pm m_z$ yields the dominant time scale. Since the effective potential scales as $L_z \sim m_z^4$ and the time scale of the dynamics is given by $\tau_z \sim |\epsilon_z|^{-1/2}$, the cost of action associated with a single flip satisfies
\begin{align} \label{eq:SDW_4}
\Delta S_z^\text{flip} \sim |m_z|^4 \tau_z \sim |\epsilon_z|^{3/2}\,,
\end{align}
which in turn implies the mean flipping time
\begin{align}
\tau^\text{flip}_z \sim \mathrm{e}^{c' N \Delta S_z^\text{flip}} \sim \mathrm{e}^{c' N |\epsilon_z|^{3/2}}
\end{align}
with a positive constant $c'$. Thus the scaling of $\Delta\Lambda$ in this regime is given by
\begin{align}
	\Delta\Lambda(\epsilon_z,N) \sim N^{k-1}\mathrm{e}^{-c' N |\epsilon_z|^{3/2}}.
\end{align}
The crossover between the above two scaling regimes is described by a scaling form
\begin{align} \label{eq:DL_fss}
\Delta\Lambda(\epsilon_z,N) = N^{k-4/3}\mathcal{G}(a\epsilon_z N^{2/3}),
\end{align}
where the asymptotic behaviors of $\mathcal{G}$ are given by
\begin{align} \label{eq:DL_fss_func}
\mathcal{G}(x) \sim \begin{cases}
 	|x|^{1/2}\mathrm{e}^{-c x^{3/2}} &\text{ for $|x| \gg 1$ and $x > 0$},\\
 	|x|^{1/2} &\text{ for $|x| \gg 1$ and $x < 0$},
 \end{cases}
\end{align}
which is consistent with Eq.~\eqref{eq:DL_fss_func_SAP} and Fig.~\hyperref[fig:psisecond]{\ref*{fig:psisecond}(b)}.

Our argument thus suggests that the finite-$N$ scaling behaviors observed numerically in the SAP in Sec.~\ref{sec:exactdiagSAP} are also valid for a broad range of models with symmetry-breaking DPTs.

\subsection{Extended scaling hypothesis for finite $T$ and $N$}
\label{ssec:fin-NT}

Combining all the scaling properties discussed in this section, we propose a joint scaling form covering the case where $N$ and $T$ are both finite. If $\mathcal{O}$ is an observable that scales as $N^y$ at criticality, and if $\langle\cdot\rangle_{\varphi(0),z}$ denotes an average over all histories constrained by the given values of $\varphi(0)$ and $z$, we propose an extended scaling hypothesis valid close to a DPT
\begin{align} \label{eq:extended_scaling}
\langle \mathcal{O} \rangle_{\varphi(0),z} = N^{y}\,\mathcal{F}(\varphi(0) N^{1/3},a \epsilon_z N^{2/3},T N^{-1/3}),
\end{align}
where $T$ in the last argument is already rescaled by Eq.~\eqref{eq:rescaling}. It is straightforward to show that the scaling forms presented above are special instances of this scaling form.
\begin{enumerate}
\item For $\mathcal{O} = \varphi(T)$, we use the scaling exponent $y = -\frac{2}{3}$, so that
\begin{align} \label{eq:phiT_extended_scaling}
\langle \varphi(T) \rangle_{\varphi(0),z} = N^{-2/3}\,\mathcal{F}(\varphi(0) N^{1/3},a \epsilon_z N^{2/3},T N^{-1/3}).
\end{align}
In the limit where $N \to \infty$ while $T$ is kept finite, let us define the reduced scaling forms
\begin{align}
\mathcal{F}_1(x) &\equiv \lim_{x'\to 0}\frac{x'^2}{x^2}\mathcal{F}\!\left(\frac{x}{x'},0,x'\right), \nonumber\\
\mathcal{F}_2(x) &\equiv \lim_{x'\to 0}\lim_{x''\to\infty}\frac{x'^2}{x}\mathcal{F}\!\left(x'',\frac{x}{x'^2},x'\right).
\end{align}
It is straightforward to show that these scaling forms satisfy
\begin{align}
	\varphi(T) &= \begin{cases}
\varphi(0)^{2} \,\mathcal{F}_1(T\,\varphi(0)) \quad \text{for $T \sim \varphi(0)^{-1}$}, \\ 	
|\epsilon_z| \,\mathcal{F}_2(T^2\,a\epsilon_z) \quad \text{for $T \sim |\epsilon_z|^{-1/2}$},
 \end{cases}
\end{align}
which reproduce the finite-$T$ scaling hypothesis shown in Eq.~\eqref{eq:phi_FSS}.
\item We may choose $\mathcal{O} = T_\text{traj}$, which denotes a dominant time scale (in the microscopic unit before the rescaling by Eq.~\eqref{eq:rescaling}) governing the evolution of the conditioned trajectory ensemble. The observable is inversely proportional to the spectral gap $\Delta\Lambda$, whose scaling exponent is $y = k-\frac{4}{3}$; thus Eq.~\eqref{eq:extended_scaling} implies
\begin{align}
\langle T_\text{traj} \rangle_{\varphi(0),z} = N^{-k+4/3}\,\mathcal{F}(\varphi(0) N^{1/3},a \epsilon_z N^{2/3},T N^{-1/3}).
\end{align}
In the limit where $T \to \infty$ while $N$ stays finite, we define a scaling form
\begin{align}
\mathcal{G}(x) \equiv \lim_{x'\to\infty}\mathcal{F}(\cdot,x,x'),
\end{align}
where the first argument of $\mathcal{F}$ can take any value due to the initial state being irrelevant as $T$ goes to infinity. Then we obtain
\begin{align}
\langle T_\text{traj}\rangle(\epsilon_z,N) = N^{-k+4/3}\mathcal{G}(a\epsilon_z N^{2/3}),
\end{align}
which is consistent with the finite-$N$ scaling hypothesis for $\Delta\Lambda$ shown in Eq.~\eqref{eq:DL_fss}.
\end{enumerate}

The extended scaling hypothesis~\eqref{eq:extended_scaling} will be useful for studying critical phenomena near a symmetry-breaking DPT observed by numerical or empirical sampling of histories, for which the system size and the observation period are both finite.

\section{Conclusions} \label{sec:conclusions}

In this paper, we introduced a class of single-box systems coupled to a pair of particle reservoirs. In the joint limit where the maximum number of particles $N$ and the observation period $T$ go to infinity, we showed analytically  that such systems exhibit symmetry-breaking dynamical phase transitions (DPTs) in the form of second-order singularities in current or activity large deviations. Although the systems are zero-dimensional, their DPTs were found to reproduce the same critical exponents as those of extended diffusive systems coupled to boundary reservoirs. In addition, for the special case of the Symmetric Antiferromagnetic Process (SAP), we numerically identified the scaling exponents governing how finite $T$ or $N$ alters the singular behaviors around a DPT. We also found theoretical explanations for these exponents, using a generic dynamical Landau theory, which imply that the same exponents apply to other single-box models in general. While our discussions focused on the cumulant generating functions defined for conditioned trajectory ensembles, it is natural to expect that these scaling exponents also govern the rounding of the conjugate large deviation functions at finite $T$ or $N$, which are more readily observable in empirical experiments.

Despite the huge difference in the number of degrees of freedom, the single-box models capture the essence of the symmetry-breaking mechanism involving the longest-wavelength mode of an extended diffusive system. Thus it seems reasonable to conjecture that the critical phenomena of these two kinds of systems belong to the same universality class --- the role played by the macroscopic length scale $L$ in an extended system should be fully equivalent to that of $N$ in a single-box model. Based on these considerations, it would be interesting to apply our finite-$N$ and finite-$T$ scaling hypotheses to identifying symmetry-breaking DPTs from the numerical or empirical data generated by extended diffusive systems.

\begin{acknowledgements}
We thank Robert L.~Jack for helpful discussion. YB is supported in part by the European Research Council under the Horizon 2020 Programme, ERC Grant Agreement No. 740269.
VL is supported by the ERC Starting Grant No. 68075 MALIG, the ANR-18-CE30-0028-01 Grant LABS and the ANR-15-CE40-0020-03 Grant LSD. YK acknowledges support from Israel Science Foundation and a US-Israel Binational-Science-Foundation grant. 
\end{acknowledgements}

\appendix

\section{Derivation of the path-integral representation}
\label{app:path}

Here we present a detailed derivation of Eqs.~\eqref{eq:cgf_path_integ} and \eqref{eq:h_before_res_0} based on the method described in \cite{LefevreJSM2007,BaekJSM2016}. First, we discretize time by dividing the observation time $[0,\,T]$ into $\mathcal{N}$ short time intervals of duration $\Delta t$, so that $T = \mathcal{N}\Delta t$. For the $s$-th time interval $t \in [(s-1)\Delta t,s\Delta t]$, we define the random variables $I^{(a)}_s$ and $I^{(b)}_s$ as
\begin{align}
	I^{(r)}_s = \begin{cases}
 1 & \text{if a particle hops to the right across a bond next to reservoir $r$,}\\
 -1 & \text{if a particle hops to the left across a bond next to reservoir $r$,}\\
 0 & \text{if nothing happens,}
 \end{cases}
\end{align}
with $r \in \{a,b\}$. Then, using the definitions of $J_T$ and $K_T$ shown in Eqs.~\eqref{eq:j_def} and \eqref{eq:k_def}, the first equation of Eq.~\eqref{eq:cgf_path_integ} can be rewritten as
\begin{align}
\rexp^{T \Psi(\lambda,\mu)} &= \left\langle \rexp^{T(\lambda J_T + \mu K_T)}\right\rangle = \left\langle \rexp^{\sum_{s=1}^\mathcal{N} \left[\frac{\lambda}{2}\left(I^{(a)}_s+I^{(b)}_s\right) + \frac{\mu}{2}\left(\left|I^{(a)}_s\right|+\left|I^{(b)}_s\right|\right)\right]}\right\rangle.
\end{align}
To convert this expression into a path integral form, we note that in the discretized dynamics the state of the box is updated according to
\begin{align}
n_s - n_{s-1} = I^{(a)}_s - I^{(b)}_s \quad \text{for $s = 1,\ldots,\mathcal{N}$},
\end{align}
where we used a shorthand notation $n_s \equiv n(s\Delta t)$. Thus we can write
\begin{align} \label{eq:exp_Psi_I_avg}
	&\rexp^{T\Psi(\lambda,\mu)} = \sum_{n_0=0}^N\cdots\sum_{n_\mathcal{N}=0}^N\, P_{n_0}\,\prod_{s=1}^\mathcal{N} 
		\left\langle\delta\!\left(n_s - n_{s-1} - I^{(a)}_s + I^{(b)}_s\right)\rexp^{\frac{\lambda}{2}\left(I^{(a)}_s+I^{(b)}_s\right) + \frac{\mu}{2}\left(\left|I^{(a)}_s\right|+\left|I^{(b)}_s\right|\right)}\right\rangle_I \nonumber\\
	&~ = \sum_{n_0=0}^N\cdots\sum_{n_\mathcal{N}=0}^N\, P_{n_0}\,\prod_{s=1}^\mathcal{N}  \int_{-i\infty}^{i\infty} \frac{d\hn_s}{2\pi}\,\rexp^{-\hn_s(n_s-n_{s-1})}
		\left\langle\rexp^{\hn_s\left(I^{(a)}_s - I^{(b)}_s\right)+\frac{\lambda}{2}\left(I^{(a)}_s+I^{(b)}_s\right) + \frac{\mu}{2}\left(\left|I^{(a)}_s\right|+\left|I^{(b)}_s\right|\right)}\right\rangle_I,
\end{align}
where $P_{n_0}$ denotes the initial state distribution, $\langle\cdot\rangle_I$ stands for the average over all possible sequences of $I^{(a)}_s$ and $I^{(b)}_s$, and the second equation is obtained by the Fourier transform of each Dirac delta function. We also note that $\hn_s$ corresponds to the auxiliary field variable in the standard Martin--Siggia--Rose (MSR) formalism~\cite{MartinPRA1973,*JanssenZPB1976,*DeDominicisJPC1976,*DeDominicisPRB1978}. The average $\langle\cdot\rangle_I$ can be evaluated using the following probability distribution of all possible outcomes
\begin{align}
	\left(I^{(a)}_s,I^{(b)}_s\right) = \begin{cases}
		(1,0) &\text{with probability $W_\rR(\bn_a,n_s)\Delta t$}, \\
		(-1,0) &\text{with probability $W_\rL(\bn_a,n_s)\Delta t$}, \\
		(0,1) &\text{with probability $W_\rR(n_s,\bn_b)\Delta t$}, \\
		(0,-1) &\text{with probability $W_\rL(n_s,\bn_b)\Delta t$}, \\
		(0,0) &\text{otherwise},
	\end{cases}
\end{align}
which simply follows from the definitions of the hopping rates. Thus we have
\begin{align}
\left\langle\rexp^{\hn_s\left(I^{(a)}_s - I^{(b)}_s\right)+\frac{\lambda}{2}\left(I^{(a)}_s+I^{(b)}_s\right) + \frac{\mu}{2}\left(\left|I^{(a)}_s\right|+\left|I^{(b)}_s\right|\right)}\right\rangle_I = 1 + \mathcal{H}_{\lambda,\mu}(n_s,\hn_s)\Delta t = \rexp^{\mathcal{H}_{\lambda,\mu}(n_s,\hn_s)\Delta t + O\!\left(\Delta t^2\right)},
\end{align}
where $\mathcal{H}_{\lambda,\mu}$ is as defined in Eq.~\eqref{eq:h_before_res_0}. Using this result in Eq.~\eqref{eq:exp_Psi_I_avg}, we find
\begin{align}
\rexp^{T\Psi(\lambda,\mu)} = \sum_{n_0=0}^N\cdots\sum_{n_\mathcal{N}=0}^N\int_{-i\infty}^{i\infty} \frac{d\hn_1}{2\pi}\cdots\int_{-i\infty}^{i\infty} \frac{d\hn_\mathcal{N}}{2\pi}\, P_{n_0}\,\rexp^{-\sum_{s=1}^\mathcal{N} \left[\hn_s(n_s-n_{s-1})-\mathcal{H}_{\lambda,\mu}(n_s,\hn_s)\Delta t\right]}.
\end{align}
Taking the limit where $\mathcal{N}$ goes to infinity, we can replace $\sum_s$ with a time integral and introduce a shorthand notation
\begin{align}
\sum_{n_0=0}^N\cdots\sum_{n_\mathcal{N}=0}^N\int_{-i\infty}^{i\infty} \frac{d\hn_1}{2\pi}\cdots\int_{-i\infty}^{i\infty} \frac{d\hn_\mathcal{N}}{2\pi}\, P_{n_0} \xrightarrow{\mathcal{N}\to\infty} \int \mathcal{D}[n,\hn].
\end{align}
Thus we finally obtain Eqs.~\eqref{eq:cgf_path_integ} and \eqref{eq:h_before_res_0}.

\section{Generalization to nonzero boundary driving}
\label{app:boundary_driving}

If the hopping rate has a multiplicative form
\begin{align}
W_\rR(n_1,n_2) = U(n_1)\,V(n_2),
\end{align}
the results discussed above can readily be generalized to the case of nonzero boundary driving $\bn_a \neq \bn_b$. In this case, the particle-hole symmetry requires
\begin{align} \label{eq:p-h_sym_bcs_bdrive}
\bn_a = N - \bn_b.
\end{align}
Using Eqs.~\eqref{eq:p-h_sym_rates}, \eqref{eq:nu}, \eqref{eq:factorize}, \eqref{eq:p-h_sym_bcs_bdrive}, and introducing shorthand notations $\bU_a \equiv U(\bn_a)$ and $\bU_b \equiv U(N-\bn_a)$, the four hopping rates in Eqs.~\eqref{eq:left_res} and \eqref{eq:right_res} can be written as
\begin{align} \label{eq:rates_simplify_APP}
W_\rR(\bn_a,n) &= \bU_a\,V(n),&
W_\rL(\bn_a,n) &= \frac{1}{\nu}\,\bU_b\,V(N-n), \nonumber\\
W_\rL(n,\bn_b) &= \frac{1}{\nu}\,\bU_b\,V(n),&
W_\rR(n,\bn_b) &= \bU_a\,V(N-n).
\end{align}
We note that, to impose the bound $0 \le n \le N$, the hopping rates are further constrained by
\begin{align} \label{eq:rate_n_constraint_bdrive}
U(0) = V(N) = 0.
\end{align}
Using these hopping rates in Eq.~\eqref{eq:h_before_res_0}, the Hamiltonian again takes the form shown in the last line of Eq.~\eqref{eq:h_before_res}, except that $\gamma \equiv (\nu\bU_a + \bU_b)/\nu$ and
\begin{align} \label{eq:z_def_bdrive}
z(\lambda,\mu) \equiv \frac{\rexp^{\mu/2}\cosh\!\left(\frac{\lambda}{2}+\tanh^{-1}\frac{\nu \bU_a-\bU_b}{\nu \bU_a+\bU_b}\right)}{\cosh\!\left(\tanh^{-1}\frac{\nu \bU_a-\bU_b}{\nu \bU_a+\bU_b}\right)}.
\end{align}
Thus the nonzero boundary driving only modifies the axis of the Gallavotti--Cohen symmetry.

\section{A note on the steady-state distribution} \label{sec:ssd}

To ensure that the symmetric profile~\eqref{eq:sym_saddle} gives the true optimal profile for $\lambda$ and $\mu$ close to zero, we also require that $\rho = 1/2$ gives the typical state of the system in the (unconditioned) steady state. To identify the criteria for this requirement, we revisit the rate equations~\eqref{eq:left_res} and \eqref{eq:right_res}. Eq.~\eqref{eq:rates_simplify_APP} implies that the rate equations can be combined into a single equation
\begin{align} \label{eq:rates_comb}
n \xrightleftharpoons[\gamma V(N-n-1)]{\gamma V(n)} n+1.
\end{align}
Thus, after the rescaling of all variables by the powers of $N$, the steady-state distribution satisfies
\begin{align}
P_\rs\!\left(\frac{1}{2}+\delta\rho\right) = P_\rs\!\left(\frac{1}{2}\right)\prod_{k=0}^{N|\delta\rho|}\frac{v(1/2+k/N)}{v(1/2-k/N)}
\end{align}
regardless of the sign of $\delta\rho$. Clearly $P_\rs$ attains the maximum at $\rho = 1/2$ if $v(\rho)$ is a monotonically decreasing function. Moreover, for small $\delta\rho$ one can write  
\begin{align}
\ln \frac{P_\rs(1/2+\delta\rho)}{P_\rs(1/2)} = N\sum_{k=0}^{Nf|\delta\rho|} \frac{1}{N}\ln\frac{v(1/2+k/N)}{v(1/2-k/N)} \simeq N\int_0^{\delta\rho} dx\,\frac{2\bv'x}{\bv} = \frac{N\bv'}{\bv}\delta\rho^2,
\end{align}
so that, given $\bv' < 0$, $P_\rs$ can be approximated by a Gaussian distribution
\begin{align}
P_\rs(\rho) \sim \exp\left[-N\frac{|\bv'|}{\bv}\left(\rho-\frac{1}{2}\right)^2\right].
\end{align}
Thus, if one observes the system in the steady state, the typical deviation of the initial state from $\rho = 1/2$ has the scale $\sqrt{\bv/(N|\bv'|)}$. This deviation plays an important role in the finite-$T$ corrections.

\bibliography{current_transition_single_box}

\end{document}